\documentclass[twocolumn,showpacs,preprintnumbers,superscriptaddress,prb,amsmath,amssymb]{revtex4-1}

\usepackage{graphicx}
\usepackage{dcolumn}
\usepackage{ulem}
\usepackage{gensymb}
\usepackage{braket}
\usepackage{epstopdf}
\usepackage{amsmath}
\usepackage{mhchem}
\usepackage{xcolor}

\begin{document}



\newcommand{\BYIO}{\ce{Ba2YIrO6}}
\newcommand{\SYIO}{\ce{Sr2YIrO6}}
\newcommand{\BIYIO}{\ce{Ba2(Y$_{1-x}$Ir$_{x}$)IrO6}}
\newcommand{\BC}{\ce{BaCl2}}

\newcommand{\cre}[2]{{#1}_{#2}^{\dagger}}
\newcommand{\ann}[2]{{#1}_{#2}^{\vphantom{\dagger}}}
\newcommand{\veck}[1]{\boldsymbol{#1}}

\newcommand{\bohrmag}{\mu_\mathrm{B}}

\title{Magnetism out of disorder in a $J=0$  compound \BYIO{}}

\author{Q. Chen}
\affiliation{Department of Physics and Astronomy, University of Tennessee, Knoxville, Tennessee 37996, USA}

\author{C. Svoboda}
\affiliation{Department of Physics, The Ohio State University, Columbus, OH 43210, USA}

\author{Q. Zheng}
\affiliation{Materials Science and Technology Division, Oak Ridge National Laboratory, Oak Ridge, Tennessee 37831, USA}

\author{B. C. Sales}
\affiliation{Materials Science and Technology Division, Oak Ridge National Laboratory, Oak Ridge, Tennessee 37831, USA}

\author{D. G. Mandrus}
\affiliation{Materials Science and Technology Division, Oak Ridge National Laboratory, Oak Ridge, Tennessee 37831, USA}\affiliation{Department of Materials Science and Engineering, University of Tennessee, Knoxville, Tennessee 37996, USA}

\author{H. D. Zhou}
\affiliation{Department of Physics and Astronomy, University of Tennessee, Knoxville, Tennessee 37996, USA}

\author{J.-S. Zhou}
\affiliation{Materials Science and Engineering Program, University of Texas at Austin, Austin, Texas 78712, USA}

\author{D. McComb}
\affiliation{Department of Materials Science and Engineering, The Ohio State University, Columbus, OH 43210, USA}

\author{M. Randeria}
\affiliation{Department of Physics, The Ohio State University, Columbus, OH 43210, USA}

\author{N. Trivedi}
\email{trivedi.15@osu.edu}
\affiliation{Department of Physics, The Ohio State University, Columbus, OH 43210, USA}

\author{J.-Q. Yan}
\email{yanj@ornl.gov}
\affiliation{Materials Science and Technology Division, Oak Ridge National Laboratory, Oak Ridge, Tennessee 37831, USA}

\date{\today}

\begin{abstract}
We systematically investigate the magnetic properties and local structure of \BYIO{} to demonstrate that Y and Ir lattice defects in the form of antiphase boundary or clusters of antisite disorder affect the magnetism observed in this $d^4$ compound.
We compare the magnetic properties and atomic imaging of (1) a slow cooled crystal, (2) a crystal quenched from 900\degree C after growth, and (3) a crystal grown using a faster cooling rate than the slow cooled one.
Atomic imaging by scanning transmission electron microscopy (STEM) shows that quenching from 900\degree C introduces antiphase boundary to the crystals, and a faster cooling rate during crystal growth leads to clusters of Y and Ir antisite disorder.
STEM study suggests the antiphase boundary region is Ir-rich with a composition of \BIYIO{}.
The magnetic measurements show that \BYIO{} crystals with clusters of antisite defects have a larger effective moment and a larger saturation moment than the slow-cooled crystals. Quenched crystals with Ir-rich antiphase boundary shows a slightly suppressed saturation moment than the slow cooled crystals, and this seems to suggest that antiphase boundary is detrimental to the moment formation.
Our DFT calculations suggest magnetic condensation is unlikely as the energy to be gained from superexchange is small compared to the spin-orbit gap.
However, once Y is replaced by Ir in the antisite disordered region, the picture of local non-magnetic singlets breaks down and magnetism can be induced. This is because of (a) enhanced interactions due to increased overlap of
orbitals between sites, and, (b) increased number of orbitals mediating the interactions.
Our work highlights the importance of lattice defects in understanding the experimentally observed magnetism in \BYIO{} and other $J=0$ systems.
\end{abstract}

\maketitle

\section{Introduction}

The magnetic properties of octahedrally coordinated Re$^{3+}$, Ru$^{4+}$, Os$^{4+}$, and Ir$^{5+}$ with a $d^4$ electronic configuration were first studied around 1960s.\cite{earnshaw1961601}
The large crystal electric field prefers the low spin configuration for these 4d/5d transition metal compounds.
The strong spin orbit coupling and the hybridization between transition metal $d$-orbital and oxygen $p$-orbital further complicate the magnetic behavior by reducing the magnetic moment or the orbital angular momentum, respectively.
In the presence of a strong spin-orbit coupling, a temperature independent magnetic susceptibility is expected.

Both $4d$ and $5d$ transition metal ions in the $d^4$ (ie. $t_{2g}^4 e_g^0$) configuration are expected to be non-magnetic in both the weakly and strongly correlated limits.
In the weakly correlated picture, $t_{2g}$ shells are split into a fully filled $j = 3/2$ shell and an empty $j=1/2$ shell due to strong spin-orbit coupling.
Spin-orbit coupling then opens up a bandgap between the $j=3/2$ and $j=1/2$ bands leading to a non-magnetic insulating ground state.
In the strongly correlated picture, the first two Hund's rules require each $d^4$ site to be in a total $S=1$ and total $L=1$.
Strong spin-orbit coupling then yields a local $J=0$ state on every ion.
Again the ground state is non-magnetic.

Interest in $d^4$ magnetism was revived after a report of long range magnetic order in Sr$_2$YIrO$_6$\cite{cao2014novel} and the proposal of a condensation mechanism for magnetism in $d^4$ Mott insulators.\cite{khaliullin2013excitonic} It was predicted that the hopping of electrons between neighboring sites would lead to an exchange interaction that could compete with the onsite spin-orbital singlet and generate a local moment. \cite{meetei2015novel}
The anomalous magnetic moments observed in \textit{A}$_2$YIrO$_6$ (\textit{A}=Sr, and Ba) double perovskites and the fragile magnetism in \textit{R}$_2$Os$_2$O$_7$ (\textit{R}=rare earth) have been the subject of intense theoretical and experimental investigations.\cite{khaliullin2013excitonic,cao2014novel,bhowal2015breakdown,pajskr2016possibility,phelan2016influence,meetei2015novel,sato2016spin,dey2016ba, ranjbar2015structural,corredor2016iridium, zhao2016fragile, zhang2016breakdown, wang2014lattice,laguna2015electronic}
A fundamental question still remains: what is the root cause of the anomalous magnetism in these $5d^4$ compounds?

There are two prominent explanations which have been applied to these $5d^4$ materials.
The first explanation is quenching of the orbital degrees of freedom through octahedral distortions.
Uniaxial distortions (ie. $\delta (\veck{L} \cdot \hat{n})^2$) split the $t_{2g}$ orbitals into an orbital singlet and an orbital doublet.
When the orbital doublet is lower in energy, the doublet is filled and the singlet is empty, and the conclusion is still a non-magnetic ground state.
However if the orbital singlet is lower in energy, it is occupied by two electrons while the other two electrons occupy the doublet.
This half filled orbital doublet then has a total spin $S=1$ leading to a magnetic ground state.
Cao et.~al.~\cite{cao2014novel} proposed this distortion mechanism was the cause of the 0.9 $\bohrmag / \mathrm{Ir}$ moments and the long range magnetic order at 1.3\,K in \SYIO{}.
However studies on Ba$_{2-x}$Sr$_x$YIrO$_6$ where the lattice distortion was tuned by varying the $x$ found the moment shows little dependence on the chemical pressure.
This leads to the conclusion that octahedral distortions are not responsible for moment formation.\cite{phelan2016influence,ranjbar2015structural}
It is also worth mentioning that Dey et.~al.~\cite{dey2016ba} and Corredor et. al.\cite{corredor2016iridium} observed no long range magnetic order in \BYIO{} and \SYIO{} crystals down to 0.4\,K despite the presence of magnetic moments.

The second explanation is a theoretical mechanism for ``excitonic'' condensation proposed by Khaliullin \cite{khaliullin2013excitonic} which is generally applicable to strongly spin-orbit coupled $d^4$ systems.
Each Ir$^{5+}$ ion is nominally in a local non-magnetic $J=0$ state with an energy gap of $\lambda / 2$ to the next lowest energy state of $J=1$ where $\lambda$ is the spin-orbit coupling strength appearing as $(\lambda / 2) \veck{L} \cdot \veck{S}$.
For very small superexchange interactions, the system remains in its unperturbed product state of non-magnetic $J=0$ singlets, but large superexchange interactions can result in a second-order phase transition to a magnetic ground state. Superexchange effectively allows a $J=1$ excitation to move between sites giving it a $k$-space dispersion $\omega(\veck{k})$.
The bandwidth of this dispersion is directly proportional to the superexchange coupling constant $t^2 / U$ where $t$ is the hopping and $U$ is the on-site Coulomb repulsion.
As the bandwidth of $\omega (\veck{k})$ increases, the energy gap between non-magnetic $J=0$ product state and excited $J=1$ states is reduced until the gap closes at a critical value of $(t^2 / U) / \lambda$ where the condensation of these magnetic excitations occurs.

Despite the observation of Curie moments in both \SYIO{} and \BYIO{}, density functional theory calculations have lead to contradictory results.
Initially, Bhowal et.~al.~\cite{bhowal2015breakdown} performed GGA+SOC+$U$ calculations within the plane wave basis to find antiferromagnetic (AFM) ground states in both \SYIO{} and \BYIO{} despite the absence of distortions in cubic \BYIO{}.
However, a later study by Pajskr et.~al.~\cite{pajskr2016possibility} rebutted this claim using dynamical mean field theory (DMFT) to obtain a non-magnetic ground state.
Furthermore, this study estimated the gap between the $J=0$ singlets and the $J=1$ triplets to be more than 250 meV which marginalizes the prospect of condensation.
We will argue on simpler grounds that this finding is qualitatively correct, and the condensation mechanism is not active in \BYIO{}.

In addition to the above routes to intrinsic magnetism, extrinsic effects remain plausible.
Dey et.~al.~\cite{dey2016ba} suggested the magnetism in \BYIO{} was due to paramagnetic impurities.
In a later study~\cite{corredor2016iridium}, they identified a Schottky anomaly in the specific heat of \SYIO{} pointing to the conclusion that paramagnetic impurities are responsible for the observed Curie susceptibilities without the onset of long range magnetic order.
In this study, we pursue this line of reasoning and examine the magnetic properties of three types of \BYIO{} single crystals with different heat treatments.
The varied heat treatments provide samples with antiphase boundary or clusters of antisite disorder, which allow us to probe the effect of lattice defects on the magnetic properties.
We find that larger effective moments and saturation moments are observed in crystals with Y/Ir antisite disorder as clusters. This correlation suggests that antisite disorder contributes to the observed moments. To investigate the plausibility of magnetic condensation in \BYIO{}, we develop a microscopic model for the condensation of $J=1$ triplet excitations in double perovskites.
Using tight-binding parameters from DFT, we show a large energy gap between the non-magnetic $J=0$ ground state and the lowest energy $J=1$ triplet excitations remains so that the condensation of magnetic excitations should not occur in \BYIO{}.
We then estimate how antisite disorder can contribute to the magnetism in the \textit{A}$_2$YIrO$_6$ double perovskites. In the
antisite disordered region, both the overlap of orbitals between sites and the number of orbitals mediating the interaction increase thus enhancing the interactions which break down the picture of local non-magnetic singlets and induce magnetism.

\section{Experimental}
\begin{figure}
 \centering
 \includegraphics [width = 0.47\textwidth] {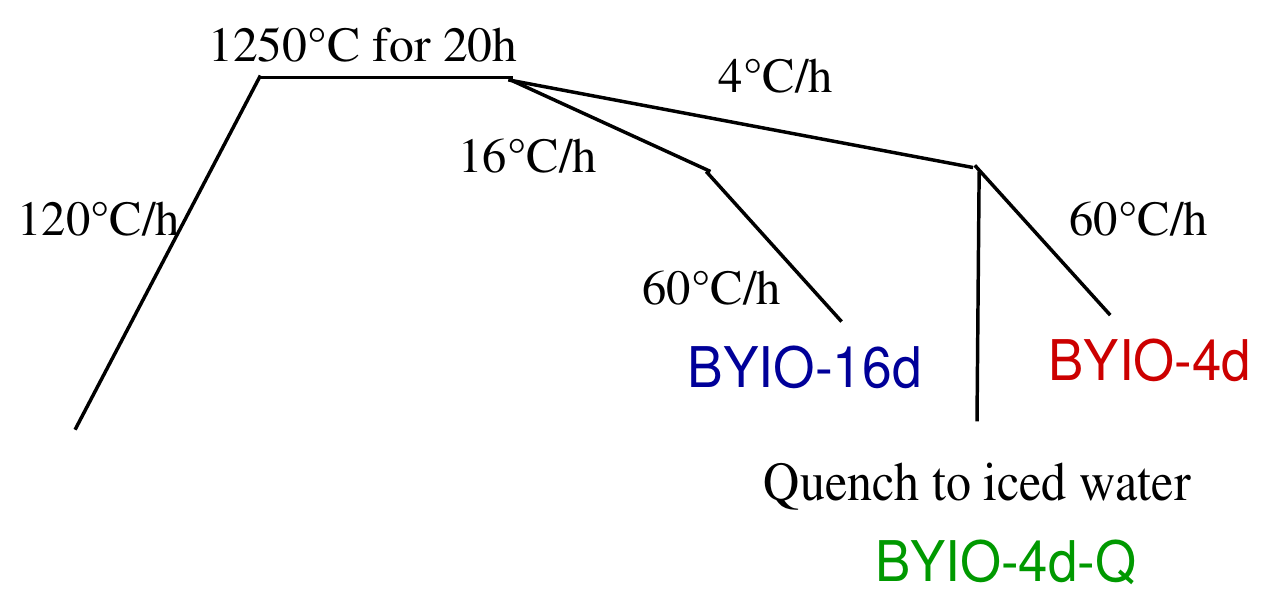}
\caption{(color online) Temperature profiles for the growth of \BYIO{} crystals. }
\label{TProfile}
\end{figure}

\BYIO{} single crystals were grown out of BaCl$_2$ flux starting with presynthesized polycrystalline \BYIO{}  materials.
The polycrystalline \BYIO{} was prepared by the conventional solid-state reaction method.
A homogeneous mixture of stoichiometric amount of starting materials \ce{BaCO3}, \ce{IrO2} and \ce{Y2O3} with purities not less than 99.9\% was pelletized and sintered  at 900\degree C for 6 hours.
The pellets were then ground into fine powder, pelletized, and fired at 1050\degree C for 15 hours.
The last sintering was performed at 1200\degree C for 30 hours with one intermediate grinding.
Room temperature x-ray powder diffraction measurement shows that \BYIO{} powder synthesized using the above procedure has about 5\% Y$_2$O$_3$ impurity.
The presynthesized \BYIO{} powder was then mixed with BaCl$_2$ in a mass ratio 1:20.
The homogeneous mixture of charge and flux was kept inside of a 20\,ml sized Pt crucible with a cover.
The crystal growth was performed using three different cooling processes as illustrated in Fig.~\ref{TProfile}.
\BYIO{} single crystals  were collected after dissolving the BaCl$_2$ flux in hot water.

Room temperature X-ray powder diffraction patterns were collected on a X'Pert PRO MPD X-ray Powder Diffractometer using the Ni-filtered Cu-K$_\alpha$ radiation. The Rietvelt refinement of the diffraction patterns was performed using Fullprof package. The elemental analysis was performed using a Hitachi TM-3000 tabletop electron microscope equipped with a Bruker Quantax 70 energy dispersive x-ray (EDX) system. The elemental analysis does not observe deviation from the desired stoichiometry. Magnetic properties were measured with a Quantum Design (QD) Magnetic Property Measurement System in the temperature interval 1.8\,K\,$\leq$\,T\,$\leq$\,750\,K. Thermopower around room temperature was measured in a homemade setup.

Scanning transmission electron microscopy (STEM) specimens were prepared by crushing \BYIO{} crystals in methanol. Drops of the resulting suspensions were deposited on lacey carbon transmission electron microscopy (TEM) grids and then dried in air. High-angle annular dark-field-STEM (HAADF-STEM) imaging was performed on two aberration-corrected STEM machines, FEI Titan S 80-300 operating at 300 kV (for BYIO-4d and BYIO-4d-Q) with a probe convergence angle of 25.2 mrad and an inner collection angle of 65 mrad, and Nion UltraSTEM 100$^\mathrm{TM}$ operating at 100 kV\cite{krivanek2008electron} (for BYIO-16d) with a probe convergence angle of 30 mrad and an inner collection angle of 86 mrad. All measurements were done at room temperature.

\section{Results}

\subsection{X-ray powder diffraction and thermopower}
With the motivation of investigating the possible effects on magnetism of lattice defects, we grew Ba$_2$YIrO$_6$ crystals using different cooling rates as shown in Fig.~\ref{TProfile}. After homogenizing at 1250\degree C, sample BYIO-4d was cooled down to 900\degree C at a cooling rate of 4\degree C per hour and then to room temperature at 60\degree C per hour. Sample BYIO-4d-Q was quenched in iced water after cooling from 1250\degree C to 900\degree C at 4\degree C per hour. Sample BYIO-16d was cooled from 1250\degree C to 900\degree C at 16\degree C per hour which is followed by cooling to room temperature at 60\degree C per hour. Room temperature x-ray powder diffraction measurements of pulversized crystals suggest the presence of Y$_2$O$_3$ in all three crystals. The Rietvelt refinement of the diffraction pattern suggests about 2\% Y$_2$O$_3$ in BYIO-4d sample. This is in line with the observation by Dey et al. The amount of Y$_2$O$_3$ increases to 6\% in BYIO-4d-Q and 11\% in BYIO-16d. Dey et al\cite{dey2016ba} proposed that the observed Y$_2$O$_3$ exists as inclusions in crystals. While this is quite likely, we occasionally observe under optical microscope isolated Y$_2$O$_3$ in BYIO-16d. The lattice parameters are a=8.3415(2)${\AA}$, 8.3438(2)${\AA}$, and 8.3416(2)${\AA}$ for BYIO-4d, BYIO-4d-Q, and BYIO-16d, respectively. The different cooling processes do not affect the lattice parameters of the as-grown \BYIO{} crystals. In the Rietvelt refinement of each pattern, we also tried to include the antisite disorder. Including about 2\% Y/Ir antisite disorder can marginally improve the quality of Rietvelt refinement. However, the refinement suggests all three crystals have the similar amount of antisite order.

One important concern is whether the different cooling processes change the stoichiometry of the resulting crystals. Our EDS measurements do not suggest, in resolution limit, that the different cooling processes lead to any deviation for Ba, Y and Ir from the desired ratio in the resulting crystals.  Dye et al. annealed the as-grown \BYIO{} crystals under 700\,bar oxygen pressure at 500\degree C for 2 days but did not notice any change to the magnetic properties.\cite{dey2016ba} This suggests that the amount of oxygen vacancies in the slow cooled crystals, if any, should be negligible and below the detection limit (around 2\%) of Thermal Gravity Analysis (TGA) or neutron diffraction. We thus measure the thermopower of our \BYIO{} single crystals. All crystals show a temperature independent thermopower about 200 $\mu$V/K around room temperature.  We also measured the electrical resistivity of dense \BYIO{} pellets quenched from 900\degree C and compared it with that of a slow cooled pellet. Both samples show a resistive behavior with a gap of 0.2 eV.
With the above crystal characterization, we would believe that the difference among those three crystals in magnetic properties should be attributed to factors other than stoichiometry, such as antisite disorder revealed by atomic mapping in this study.

\subsection{STEM}
In the perfect Ba$_2$YIrO$_6$ structure, four crystallographic sites are expected to be fully occupied by Ba, Y, Ir, and O, respectively. In fully ordered state, Y and Ir ions order into the rock-salt arrangement. However, this ordered arrangment can be destroyed at antiphase boundary or by antisite disorder where Y cation occupies the position of Ir or vice versa. Heat treatment is an effective approach to tune the degree of the cation order of double perovskites in materials processing. It is expected that the faster cooling during crystal growth and quenching from high temperature might introduce defects such as antisite disorder or antiphase boundary. In order to investigate possible lattice defects induced by quenching from high temperature or fast cooling during crystal growth, we performed STEM on all three samples.

\begin{figure*}[t]
\centering
\includegraphics[width=0.80\textwidth]{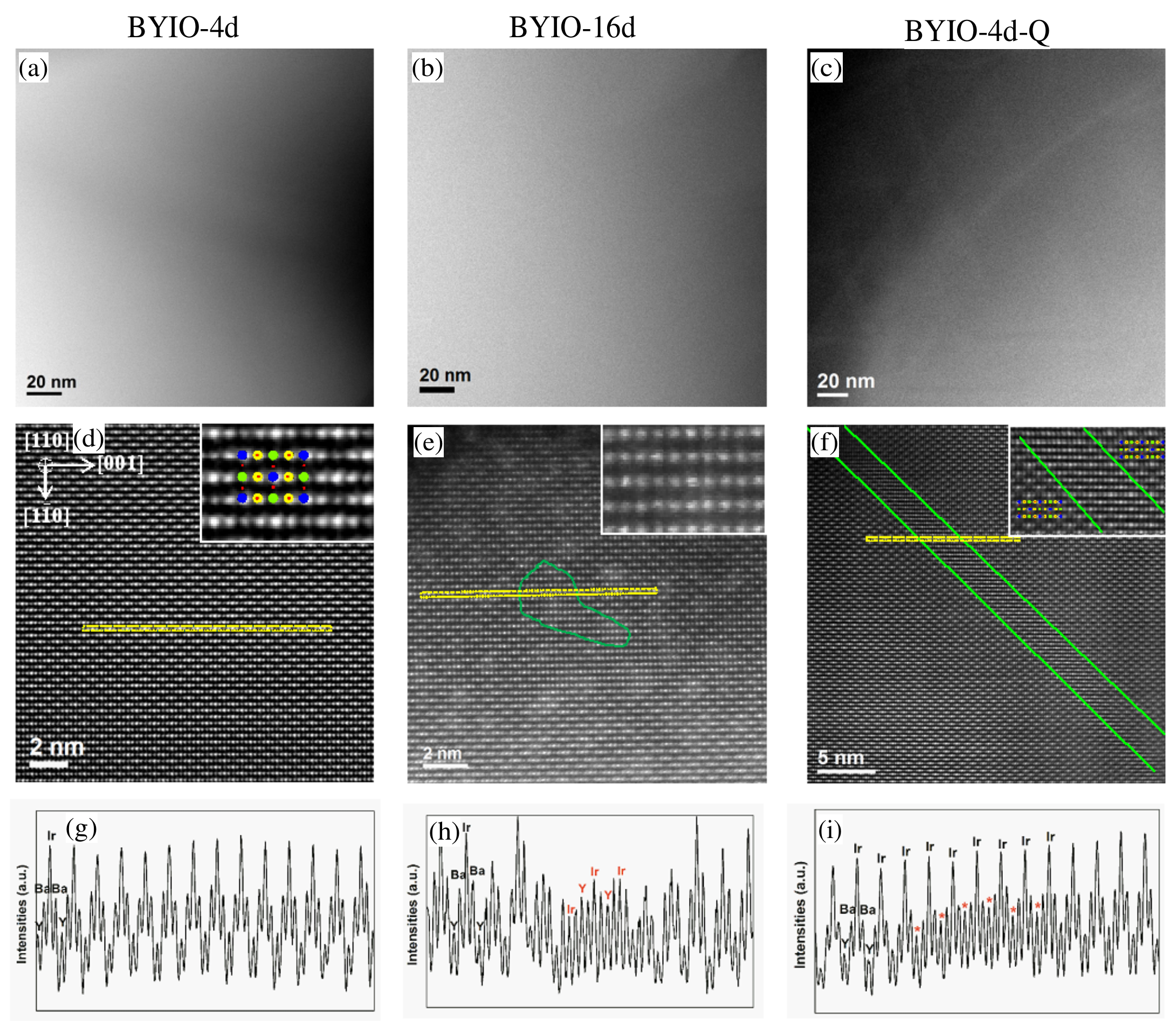}
\caption{(Color online) HADDF-STEM images in the [110] projection of \BYIO{} crystals. (a, b, c) show the low magnification HAADF image for BYIO-4d, BYIO-16d, and BYIO-4d-Q, respectively. (d, e, f) show the high resolution HAADF image for BYIO-4d-NQ, BYIO-16d, and BYIO-4d-Q, respectively. Inset of (d) highlights the ordering in BYIO-4d (yellow: Ba, green: Y, blue: Ir, and red: O). Inset of (e) shows the details of a disordered region of BYIO-16d. Inset of (f) highlights a narrow defected region through the ordered crystal.(g, h, i) show the intensity profile of atomic columns along [001] as marked by the (yellow) lines in (d, e, and f) for BYIO-4d, BYIO-16d, and BYIO-4d-Q, respectively. }
\label{STEM}
\end{figure*}

Along [001], Y and Ir atoms are always projected in the same columns. However, three types of cationic columns could be sequentially imaged in the [110] projection with an arrangement of Ba-Y-Ba-Ir-Ba columns. That means imaging along [110] can better reveal atomic-level distribution of the cations, especially the ordering behavior of cations at Y and Ir sites. As shown below, in the [110] projection, Ba, Y, and Ir atoms can be well distinguished by Z contrast. Therefore, STEM is an ideal tool to resolve the microstructural defects that might be related to the observed magnetic moments in \BYIO{}.

Figure~\ref{STEM}(a-f) show the HAADF images along [110] for all three \BYIO{} crystals.
The low magnification HAADF image for BYIO-4d crystal (see Fig.~\ref{STEM}(a)) reveals it is chemically homogenous without any large scale chemical inhomogeneity or structural disorder.
This homogeneity is also evidenced by its high-resolution HAADF (HR-HAADF) image as shown in Fig.~\ref{STEM}(d).
A typical intensity profile of columns (Fig.~\ref{STEM}(g)) in this image shows ideal sequential arrangement of Ba-Y-Ba-Ir-Ba columns, and no structural defects are found within the resolution limit of HAADF imaging.

This structural perfection is destroyed when the \BYIO{} crystal is grown using a faster cooling rate during crystal growth.
While the low magnification HAADF image for BYIO-16d crystal (Fig.~\ref{STEM}(b)) does not show any chemical or structural disorder in a large scale, a careful investigation of the high resolution (HR) HAADF image find disordered regions as highlighted by the (green) curves in Fig.~\ref{STEM}(e).
The intensity profile of columns (Fig.~\ref{STEM}(h)) across the disordered region reveals strengthened intensities of Y columns  while weakened ones of Ir columns, indicating the appearance of antisite disorder.
The largest in-plane dimension of the disordered region can be around 8\,nm.

Figure~\ref{STEM}(c) shows the low magnification HAADF image for the BYIO-4d-Q crystal quenched from 900\degree C.
Different from the featureless images in Figs.\,\ref{STEM}(a) and \ref{STEM}(b), a narrow stripe-like region longer than 100\,nm can be well observed.
As revealed by HR-HAADF image shown in Fig.~\ref{STEM}(f), the width of the stripe-like region is around 5 nm.
Fig.~\ref{STEM}(i) shows the intensity profile across this defected region.
The intensity of Y columns becomes stronger in the defected region, while no obvious anomaly is observed for the intensity of Ir columns.
This is quite different from that in BYIO-16d where the increased intensity of Y columns is accompanied with the weakened intensity of Ir columns.
The difference signals that quenching from 900\degree C induces one type of defects different from antisite disorder.
The dimension of the defected region and the intensity change of Y and Ir columns suggest the formation of an antiphase boundary in the quenched sample BYIO-4Q.
The antiphase boundary seems to be Ir-rich with a composition of \BIYIO{}.
Considering possible probe channeling effect,\cite{esser2016quantitative} we cannot rule out the possible occupation of Y at Ir site in the antiphase boundary region.
However, the population of Y at Ir site, if any, should be small.

Our HAADF-STEM imaging shows that a faster cooling during crystal growth leads to antisite disordered regions with the largest dimension of 8\,nm.
While quenching from 900\degree C results in the formation of antiphase boundary of 5\,nm wide and over 100\,nm long which is Ir-rich.
Although in the BYIO-4d crystals, HAADF images show no obvious sign of antisite disorder, we could still reasonably speculate that small disordered regions, such as single site antisite disorder, exist in these crystals, which are however out of the detection limit of HAADF imaging.

\begin{figure}
\centering
\includegraphics[width=0.47\textwidth]{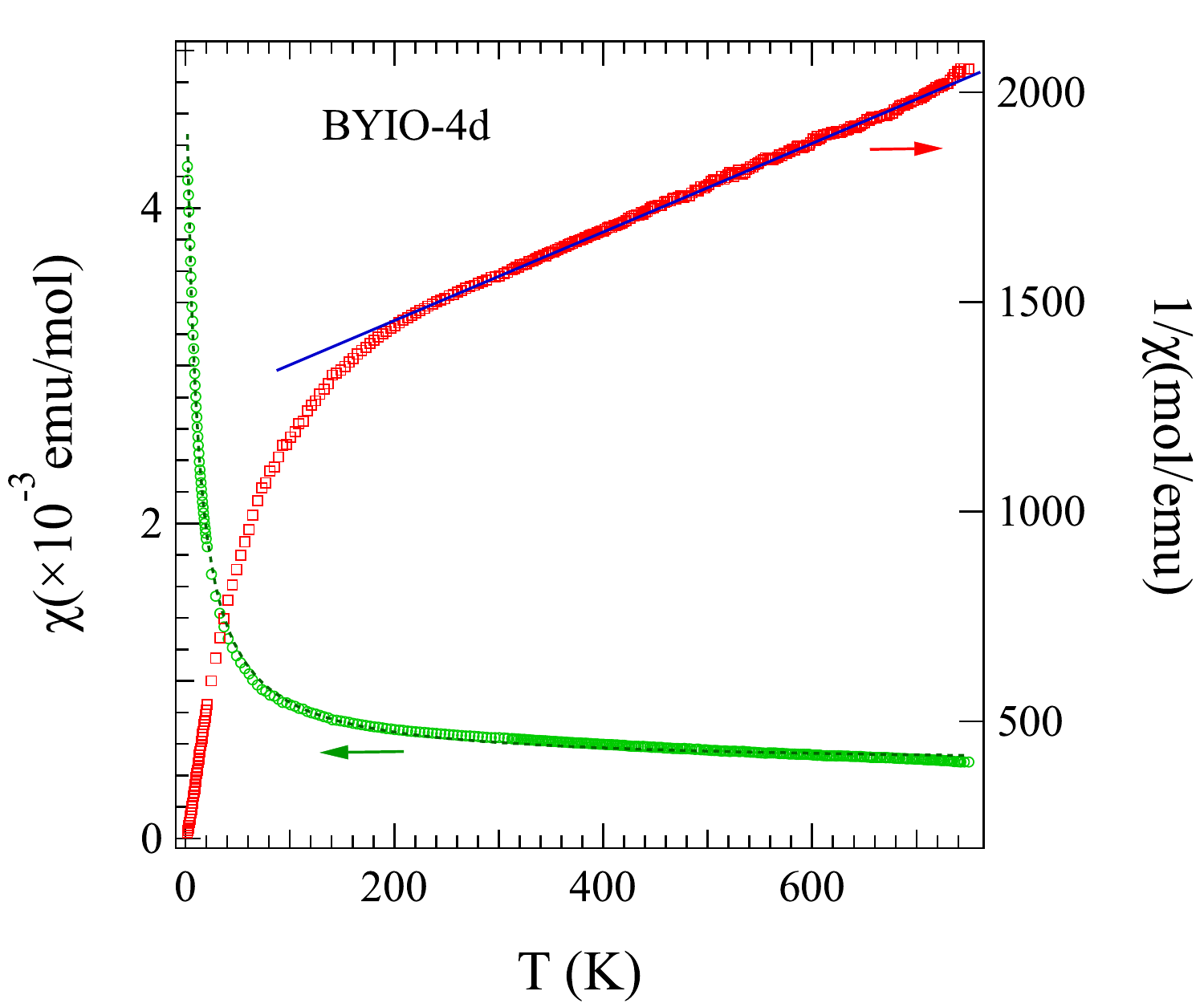}
\caption{
(color online) Temperature dependence of magnetic susceptibility $\chi$ and $1/\chi$ for BYIO-4d single crystals in the temperature range $2\,\mathrm{K} \leq T \leq 750\, \mathrm{K}$ measured in an applied magnetic field of 20\,kOe.
The solid (blue) line shows the linear fitting of 1/$\chi$ above 200\,K. The dashed curve shows the fitting with a Curie-Weiss law as described in the text.
}
\label{Chi4d}
\end{figure}

\subsection{Magnetic properties}
Figure~\ref{Chi4d} shows the temperature dependence of magnetic susceptibility of BYIO-4d measured in an applied magnetic field of 20\,kOe in both field-cooling (FC) and zero-field-cooling (ZFC) modes.
FC and ZFC curves overlap in the whole temperature range $2\,\mathrm{K} \leq T \leq 750\, \mathrm{K}$.
As reported before, no sign of long range order is observed.\cite{dey2016ba}
The magnetic susceptibility decreases with increasing temperature up to 750\,K.
This suggests the contribution from the magnetic excited states is small.
The susceptibility of a system with a nonmagnetic ground state and a magnetic excited state can be described by\cite{yan2004bond}
 \begin{equation}
\chi(T)=\frac{N_Ag^2\mu _B^2\nu S(S+1)(2S+1)e^{(-\Delta E/T)}}{3k_BT[1+\nu (2S+1)e^{-\Delta E/T}]}
 \end{equation}
where N$_A$ is the Avogadro number, g is the spectroscopic splitting factor, $\mu_\mathrm{B}$ is the Bohr magneton, $S$ is the total spin of the excited states, $\nu$ is the orbital degeneracy of the excited states, $k_B$ is the Boltzmann constant, $\Delta E$ is the energy difference between the nonmagnetic ground state and the magnetic excited state.
With $\Delta E=350\,\mathrm{meV}$,\cite{RIXS} the magnetic susceptibility from the $J=1$ state is estimated to be $9\times10^{-7}$ emu/mol and  has little effect on the temperature dependence of total magnetic susceptibility.

Figure~\ref{Chi4d} also shows the reciprocal of magnetic susceptibility.
As highlighted by the solid line, a linear temperature dependence of $1/\chi$ can be found in the temperature range $240\,\mathrm{K} \leq T \leq 750\,\mathrm{K}$.
The linear fitting using $1/\chi(T)=(T-\theta)/C$ where $C$ is the Curie constant and $\theta$ the Weiss constant, gives an effective moment $\mu_\mathrm{eff}=2.90$\,$\mu_\mathrm{B}$/Ir and a Weiss constant of about $-1200$\,K.
Both $\mu_\mathrm{eff}$ and $\theta$ are unreasonably large in a system that is nominally comprised of J=0 singlets.
As shown by the dashed curve in Fig.~\ref{Chi4d}, the temperature dependence of $\chi(T)$ can be roughly described by adding a temperature independent term $\chi_0$ in the Curie-Weiss fitting $\chi(T)=\chi_0+C/(T-\theta)$.
However, a large $\chi_0$ about $5\times 10^{-4}$ emu/mol is needed.
This value is comparable to that reported by Dey et al.\cite{dey2016ba} Before we present the Curie-Weiss fitting results of all samples, we first evaluate whether such a large $\chi_0$ is reasonable.

\begin{figure} \centering \includegraphics [width = 0.47\textwidth] {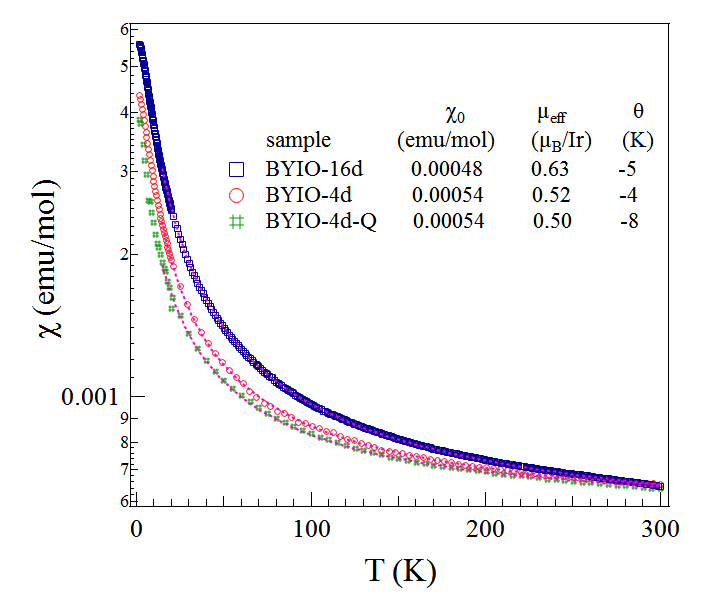}
	\caption{(color online) Temperature dependence of magnetic susceptibility of \BYIO{} single crystals in the temperature range $2\,\mathrm{K} \leq T \leq 300\,\mathrm{K}$ measured in an applied magnetic field of 50\,kOe. The dashed curves show the Curie-Weiss fitting in the temperature range $15\,\mathrm{K} \leq T \leq 300\,\mathrm{K}$. The fitting parameters are listed in the inset.}
	\label{chiT}
\end{figure}

Three contributions to $\chi_0$ should be considered for \BYIO{}: the diamagnetic susceptibility from core electrons, $\chi_\mathrm{core}$, the diamagnetic signal from the sample holder, and the Van Vleck paramagnetic susceptibility, $\chi_\mathrm{vv}$.
By adding the core electron diamagnetic susceptibility of each ion, $\chi_\mathrm{core}$ is found to be $-1.68 \times 10^{-4}$ emu/mol.\cite{dey2016ba}
The diamagnetic signal from the sample holder (straw and capsule) is estimated to be about $-1 \times 10^{-4}$ emu/mol.
For the temperature independent $\chi_\mathrm{vv}$, a simple calculation by standard perturbation theory gives
\begin{equation}
\chi_\mathrm{vv} = 2N_A\frac{{\lvert\langle{j=1}|\mu_z|j=0\rangle\rvert}^2}{\Delta E}
\end{equation}
where $\Delta E = 350$ meV according to RIXS data.\cite{RIXS}
In the $t_{2g}^4$ configuration, the magnetic moment operator can be written as $\veck{\mu} = -i\sqrt{6}(\veck{T}-\veck{T}^\dagger - g_J\veck{J})$, where $g_J = 1/2$. Thus we obtain that\cite{khaliullin2013excitonic}

\begin{equation}
\chi_\mathrm{vv} = \frac{12N_A\mu_\mathrm{B}^2}{\Delta E} = 11 \times 10^{-4} \mathrm{emu/mol}
\end{equation}

Therefore, the $\chi_0$ is estimated to be around $8 \times 10^{-4}$ emu/mol, which is larger than the fitting value of 5 $\times 10^{-4}$ emu/mol.
Figure~\ref{chiT} shows the temperature dependence of magnetic susceptibility of all three \BYIO{} crystals below room temperature.
No sign of long range magnetic order was observed above 2\,K.
We fit the data in the temperature range $15\,\mathrm{K} \leq T \leq 300\,\mathrm{K}$ using $\chi(T) = \chi_0 + C/(T-\theta)$.
The temperature range $15\,\mathrm{K} \leq T \leq 300\,\mathrm{K}$ was selected purposely for a direct comparison with the report by Dey et al.
The fitting parameters are also listed in Fig.~\ref{chiT}.
The $\chi_0$ is similar for all three samples.
BYIO-4d-Q has the smallest effective moment but the largest Weiss constant.
The effective moment for BYIO-16d is over 20\% larger than those for BYIO-4d and BYIO-4d-Q.

\begin{figure} \centering \includegraphics [width = 0.47\textwidth] {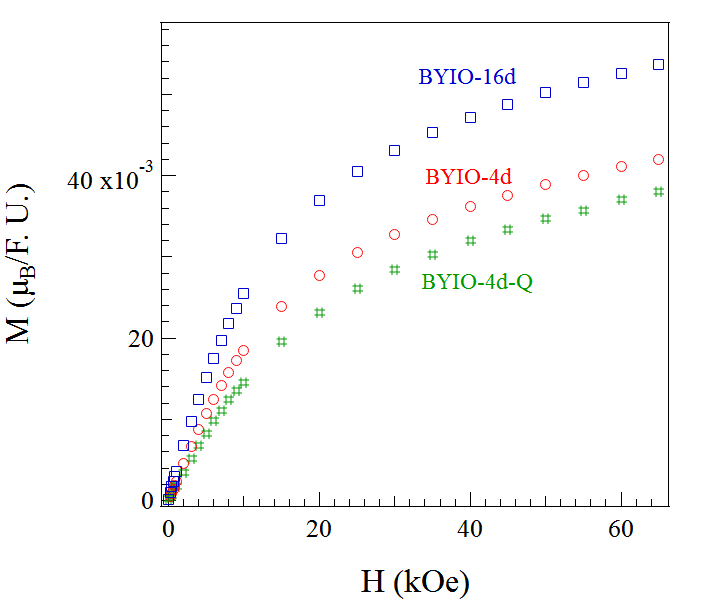}
	\caption{(color online) Field dependence of magnetization of \BYIO{} single crystals at 2\,K. }
	\label{MH}
\end{figure}

Figure~\ref{MH} shows the field dependence of magnetization measured at 2\,K. Linear fits to the magnetization versus applied magnetic field data,  $M(H)$, for the field range $H=30$-$65$ kOe gives the intrinsic susceptibility from the slope and the saturation magnetization  from the $H=0$ intercept. The intrinsic susceptibility for these three crystals is close to each other, while the saturation magnetization for BYIO-16d is much larger than that for BYIO-4d and BYIO-4d-Q. The saturation magnetization is 0.036, 0.026, and 0.022\,$\mu_\mathrm{B}$/Ir for BYIO-16d, BYIO-4d, BYIO-4d-Q, respectively.

\section{Theory}
\subsection{Condensation Mechanism}
We revisit the DFT-DMFT theory results\cite{pajskr2016possibility} that confirm a non-magnetic ground state. To gain insight into why magnetic condensation is absent, we must connect those results with the condensation mechanism which pits superexchange with an energy scale of $t^2 / U$ against the singlet-triplet energy gap of $\Delta_\mathrm{SOC}$ from spin orbit coupling.
The critical ratio of superexchange to spin-orbit coupling required to produce magnetic condensation in a single perovskite\cite{khaliullin2013excitonic} is given approximately by $10 t^2/U \ge \Delta_\mathrm{SOC}$.
In this section, we will repeat this analysis for double perovskites and find the criteria for closing the singlet-triplet gap.

\begin{figure}
\centering
\includegraphics[width=0.47\textwidth]{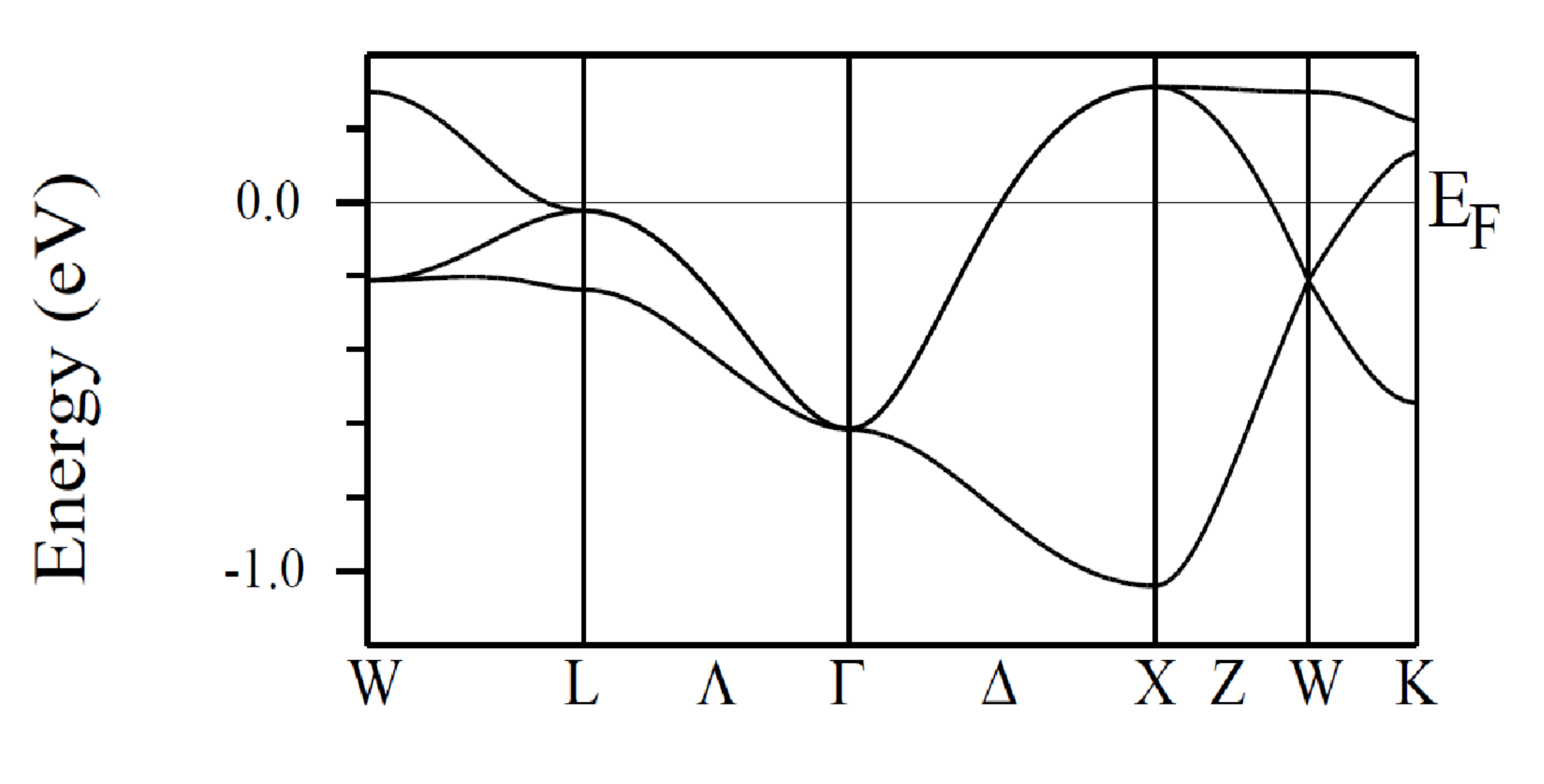}
\caption{
The GGA band structure \textit{without} spin-orbit coupling is shown.
A tight-binding model with $t_{2g}$ Wannier orbitals is fit to the three bands pictured.
\label{band}
}
\end{figure}

We first obtain a tight-binding model relevant for Y-Ir double perovskites.
The large crystal field splitting induced by $5d$ oxygen octahedral complexes separates the $e_g$ orbitals in energy so that only the $t_{2g}$ orbitals are relevant.
Then we can write our tight-binding model as the sum of electron hopping between $t_{2g}$ orbitals in the $yz$, $xz$, and $xy$ planes as $H = H_{yz} + H_{xz} + H_{xy}$.
For the $xy$ plane, for example, we can restrict the form of the tight-binding model by symmetry
\begin{equation}
H_{xy} =
\sum_{\langle ij \rangle \, \in \, xy}
\;
\sum_{\alpha \beta}
\,
\sum_{ \sigma \in \lbrace \uparrow,\downarrow \rbrace  }
t_{\alpha \beta} \,
\cre{c}{i\alpha\sigma}
\ann{c}{j\beta\sigma}
+
\mathrm{h.c.}
\label{tightBindingNN}
\end{equation}
\begin{equation}
t_{\alpha \beta} =  \left[ \begin{array}{ccc}
 t_{11} & t_{12} & 0 \\
 t_{12} & t_{11} & 0 \\
 0 & 0 & t_{33}
\end{array} \right]_{\alpha \beta}
\end{equation}
where $\alpha$ and $\beta$ index the $t_{2g}$ orbitals $yz$, $xz$, and $xy$ in this order and $\langle ij \rangle$ ranges over nearest neighbors within an $xy$ plane of the lattice of Ir ions.
Next-nearest neighbor hopping is ignored since the resulting superexchange constants will be negligible.
(See Appendix \ref{appendixNNNJustification}.)
We then use a DFT calculation\cite{blaha2001wien2k} to obtain these tight-binding parameters \textit{without} spin-orbit coupling to separate the energy scales for superexchange and spin-orbit coupling.
Fig.\ref{band} shows the band structure of the three ($t_{2g}$) bands.

A tight-binding fit with maximally localized Wannier orbitals\cite{mostofi2014updated} yields the following parameters: $t_{11} = +23$ meV, $t_{12} = \pm 19$ meV, and $t_{33} = -131$ meV.
Note the value of $t_{12}$ takes a different sign for different pairs of nearest neighbors due to relative orientations of $t_{2g}$ orbitals.

We now calculate the triplet excitation Hamiltonian where triplet excitations from the non-magnetic $J=0$ to excited $J=1$ states are described by $\ket{J_i=1,J_{i,z} = m} = \cre{T}{i,m} \ket{J_i = 0}$.
After performing a unitary transformation into cubic coordinates $T_x = \tfrac{1}{i\sqrt{2}} (T_1 - T_{-1})$, $T_y = \tfrac{1}{\sqrt{2}} (T_1 + T_{-1})$, and $T_z = iT_0$, the quadratic part of the effective triplet Hamiltonian has the following form
\begin{equation}
\begin{split}
 H_{xy}' = & \sum_{\langle ij \rangle \in xy} \left[
\left(
\cre{\veck{T}}{i} \cdot \boldsymbol{A} \cdot \ann{\veck{T}}{j} + \cre{\veck{T}}{i} \cdot \boldsymbol{B} \cdot \cre{\veck{T}}{j} + \mathrm{h.c.} \right) \right.
\\
 & \left.
+ \left( \cre{\veck{T}}{i} \cdot \boldsymbol{C} \cdot \ann{\veck{T}}{i} + (i \leftrightarrow j) \right)
\right]
\end{split}
\label{equationTripletXYHamiltonian}
\end{equation}
where $\cre{\veck{T}}{i} = ( \cre{T}{i,x}, \cre{T}{i,y}, \cre{T}{i,z} )$ and the matrices $\boldsymbol{A}$, $\boldsymbol{B}$, and $\boldsymbol{C}$ involve combinations of the hopping terms with an overall energy scale proportional to $t^2/U$; detailed expressions for $\boldsymbol{A}$, $\boldsymbol{B}$, and $\boldsymbol{C}$ are given in the Appendix \ref{appendixABCMatrices}.
The triplet Hamiltonian for the other two planes can be obtained from cyclic permutations on $H_{xy}'$.
The total Hamiltonian for the triplet excitations is given by the sum over all three planes and the singlet-triplet gap of $\Delta_\mathrm{SOC}$.
\begin{equation}
H' = H_{yz}' + H_{zx}' + H_{xy}' + \Delta_\mathrm{SOC}\sum_{i} \cre{\veck{T}}{i} \cdot \ann{\veck{T}}{i}
\label{fullTripletSpectrum}
\end{equation}

To determine when condensation occurs, we diagonalize the effective Hamiltonian $H'$ and obtain the energy dispersion $\omega(\veck{k})$ for triplet excitations.
Before performing the rigorous calculation with all parameters included, it is useful to obtain a simple estimate for the energy scales required to produce magnetic condensation.
To obtain this estimate, ignore the superexchange dependent contributions to the singlet-triplet gap (ie. set $\boldsymbol{C} = 0$) and only include the largest tight-binding parameter $t_{33}$ (ie. set $t_{11} = t_{12} = 0$).
In this simplified scenario, the $\boldsymbol{A}$ and $\boldsymbol{B}$ matrices are diagonal with $\boldsymbol{A} = -\boldsymbol{B}$, and a closed form solution can easily be obtained.
The condition for closing of the singlet-triplet gap is given by the following.
\begin{equation}
\frac{16}{3} \frac{t_{33}^2}{U} \ge \Delta_\mathrm{SOC}
\label{condensationCriteria}
\end{equation}
This value is approximately half of the estimate for single perovskites\cite{khaliullin2013excitonic} ($10 t^2 / U \ge \Delta_\mathrm{SOC}$) with the difference due to fewer superexchange paths available in face-centered cubic geometry compared to simple cubic geometry.
With a $U$ value taken to be $2$ eV, the left-hand side is estimated to be $45$ meV compared to the spin-orbit gap of $\Delta_\mathrm{SOC} \approx 350$ meV for \BYIO{}.\cite{RIXS}
From this estimate, the condensation mechanism should be inactive in \BYIO{}.
(Note that equation \eqref{condensationCriteria} only gives the criteria for a gap closing. The difference between the two sides of equation \eqref{condensationCriteria} does \textit{not} give the value of the remaining gap.)

\begin{figure}
\centering
\includegraphics[width=0.47\textwidth]{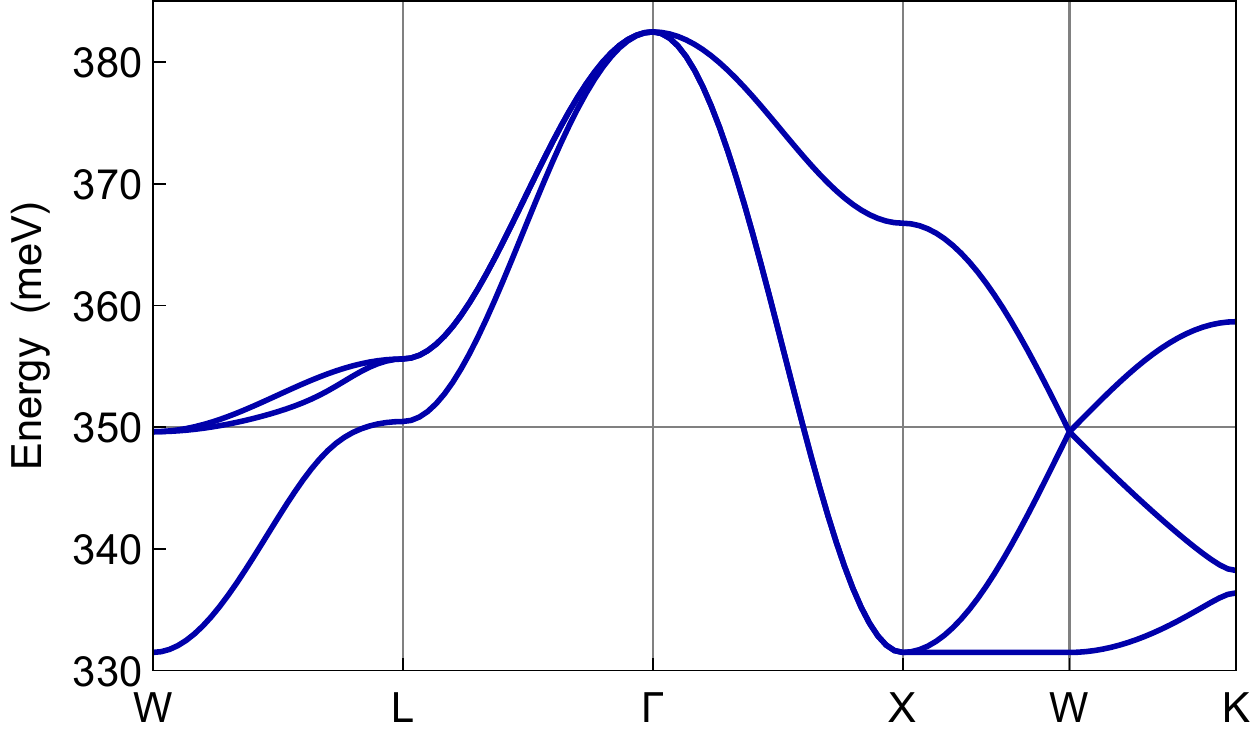}
\caption{
The triplet excitation spectrum is plotted using a spin-orbit gap of 350 meV.
Superexchange reduces the cost to make a triplet excitation at the edge of the Brillouin zone.
The lowest energy excitations appear along X-W about 20 meV lower than the original 350 meV spin-orbit gap.
}
\label{triplon}
\end{figure}

The triplet dispersion for the full model with all tight-binding parameters is shown in Fig.\ref{triplon}.
The lowest energy triplet excitations occurs along X-W approximately 20 meV below the spin-orbit gap of 350 meV.
Although superexchange interactions do reduce the energy cost for magnetic condensation, this 20 meV reduction in energy is very small compared to the overall spin-orbit gap of 350 meV.
Consequently magnetic condensation is extremely unfavorable.

\subsection{Magnetism from antisite disorder}

Since the absence of magnetism in $d^4$ Mott insulators can be understood at the atomic level, it would seem that antisite disorder would not alter the picture.
While the effects of interactions between B' sites in A$_2$BB'O$_6$ is unlikely to generate a magnetic state, the situation is different when some B sites also contain $d^4$ ions.
First, the tight-binding values ($t$) become much larger due to the increased overlap of orbitals between sites.
Hence the superexchange interaction strength, $J_\mathrm{SE} \sim t^2 / U$, becomes significantly larger.
This changes the relative strength of $J_\mathrm{SE} / \lambda$ which determines if the condensation of magnetic excitations becomes energetically favorable.
Second, the orbital geometry now consists of corner sharing octahedra with two active orbitals along a bond in constrast to the FCC lattice where only one orbital primarily contributes to superexchange through $t_{33}$.
With twice as many orbitals, there are four times as many superexchange processes involved, and hence magnetic interactions become much stronger.
With both of these contributions, the picture of local non-magnetic singlets will likely break down when both B and B' site contain Ir$^{5+}$ ions.
In this situation, the Curie moment from bulk susceptibility scales with the number of B sites containing Ir$^{5+}$ ions.

In the limit that these misplaced ions are isolated antisite defects (with no clustering of misplaced ions), the total susceptibility is then proportional to the number of antisite defects.
\begin{equation}
\chi \propto \frac{j(j+1) g^2 \mu_B^2}{3k_B (T-\theta)} \times f_\mathrm{misplaced}
\end{equation}
Here $f_\mathrm{misplaced}$ is the fraction of Ir$^{5+}$ ions appearing on B sites.
As a rough estimate, assume that for each misplaced ion, the misplaced ion and its six surrounding B' neighbors become magnetic $j=1$ states with a corresponding $g=1/2$.
Then the effective Curie moment from bulk susceptibility is given as $\mu_\mathrm{expt}^2 = \tfrac{7}{2} \mu_B^2 \times  f_\mathrm{misplaced}$.
In this scenario, even when just 1 percent of the sites were disordered, the measured bulk $\mu_\mathrm{eff}$ would be 0.18$\mu_B$.
Although this estimate serves as an upper bound, it shows that the effects of antisite disorder can amount to a substantial contribution.

\section{Discussion}
There are two different reports on the physical properties of \BYIO{} single crystals. Zhang et al. concluded that their \BYIO{} single crystals order antiferromagnetically below 1.6\,K by measuring specific heat and magnetic properties.\cite{zhang2016breakdown} In the paramagnetic state, the Curie-Weiss fitting of the magnetic susceptibility in the temperature range $50\,\mathrm{K} \leq T \leq 300\,\mathrm{K}$ gives an effective moment of 1.44$\mu_\mathrm{B}$/Ir and a Weiss constant of $-149$\,K. In contrast, Dey et al. did not observe any long range magnetic order for temperatures down to 0.4\,K in their crystals.\cite{dey2016ba} The Curie-Weiss fitting of the magnetic susceptibility in the temperature range $15\,\mathrm{K} \leq T \leq 300\,\mathrm{K}$ gives an effective moment of 0.44$\mu_\mathrm{B}$/Ir and a Weiss constant of $-8.9$\,K. Both values are much smaller than those reported by Zhang et al. Our \BYIO{} crystals are similar to what Dey et al. obtained with similar $\chi_0$ and $\theta$.

The Curie-Weiss fitting of magnetic susceptibility shown in Fig.~\ref{chiT} gives a $\chi_0$ around 5$\times10^{-4}$ emu/mol. This value agrees well with $\chi_0 = 5.83 \times10^{-4}$ emu/mol reported by Dey et al.\cite{dey2016ba} However, all our \BYIO{} crystal have an effective moment larger than 0.44\,$\mu_\mathrm{B}$/Ir reported by Dey et al.
Dey et al. did not report the cooling rate during crystal growth. As discussed later, both the cooling rates during and after crystal growth affect the magnetic properties of \BYIO{} crystals. The different growth parameters might account for the slight discrepancy in the magnitude of effective moments.

As shown in Figs. \ref{chiT} and \ref{MH}, BYIO-16d has a larger effective moment and saturation moment than BYIO-4d and BYIO-4d-Q. The magnetic moment for BYIO-4d-Q is slightly smaller than that of BYIO-4d. The crystal characterization presented in Sec. III-A suggests that the lattice defects should be an important, if not the only, factor affecting the magnetic properties of \BYIO{} crystals cooled differently. The observation of clusters of antisite disorder (see Fig.~\ref{STEM}(e)) and a larger magnetic moment (see Figs. \ref{chiT} and \ref{MH}) in BYIO-16d crystals suggests that antisite disorder favors the formation of magnetic moments out of the $J=0$ ground state. This is supported by our theoretical analysis. Inside of the antisite disordered clusters, the interactions are enhanced by increased overlap of orbitals between sites and increased number of orbitals mediating. The enhanced magnetic interactions, if strong enough, can break down the picture of local non-magnetic singlets and thus induce condensation of magnetic excitations.

The moment induced by antisite disorder is unexpected. For double perovskites with both B and B' cations magnetic, there are  ample studies showing the detrimental effect of antisite disorder on the magnetic properties. It has been demonstrated that antisite disorder suppresses the ordered moment, ordering temperature, or coervicity.\cite{lim2016insights,navarro2001cationic} For double perovskites which contain only one magnetic cation but show long range magnetic order, we would expect similar effect since antisite disorder would disturb the magnetic interactions.  From this point of view, \BYIO{} provides a special case where antisite disorder is beneficial to magnetic moments. This unique feature is rooted in the importance role of interactions in the moment formation of $d^4$ systems. The occupation of Y site by Ir enhances the interactions thus leading to the appearance of finite-spin Ir ions in their localization volume.

It is interesting to notice that the BYIO-4d-Q sample, in which stripe-like antiphase boundary up to 100\,nm is observed in atomic mapping, has a slightly smaller effective moment than the slow cooled BYIO-4d.
This suggests that the formation of antiphase boundary reduces the magnetic moment of \BYIO{}.
As shown in Fig.~\ref{STEM}(i), the intensity of Ir columns is unchanged while that of Y increases in the antiphase boundary region.
The intensity change signals some occupation of Ir at Y site forming an Ir-rich region with a composition of \BIYIO{}.
One important result of this is the appearance of magnetic Ir$^{4+}$ ions with five electrons which are expected to enhance the magnetic moment of \BYIO{}.
This is opposite to the experimental results that BYIO-4d-Q has smaller saturation and effective moments than the slow-cooled BYIO-4d.
The intensity profile shown in Fig.~\ref{STEM}(i) still suggests an atomic arrangement of Ba-Y-Ba-Ir-Ba.
We thus look at the magnetic properties of high pressure cubic phase of Ba$_3$YIr$_2$O$_9$ in order to understand the slightly reduced moment in BYIO-4d-Q.
Ba$_3$YIr$_2$O$_9$ can be stabilized in a cubic (space group \textit{Fm}-3\textit{m}) structure with an ordered arrangement of IrO$_6$ and Y$_{2/3}$Ir$_{1/3}$O$_6$ octahedra forming the cubic double perovskite Ba$_2$Ir(Y$_{2/3}$Ir$_{1/3}$)O$_6$ structure.\cite{dey2013possible}
No long range magnetic order was observed above 2\,K for the high pressure cubic phase.
The effective moment is reported to be 0.19$\mu_\mathrm{B}$/Ir, much smaller than 0.50$\mu_\mathrm{B}$/Ir for BYIO-4d-Q.
According to the first-principles study by Panda et al., the magnetic exchange interactions is weak in the cubic Ba$_3$YIr$_2$O$_9$ and spin-orbit coupling dominates leading to a quantum spin-orbit liquid state.\cite{panda2015electronic}

Our experimental study and theoretical analysis support the contribution of antisite disorder to the enhanced moment in BYIO-16d.
While the Ir-rich antiphase boundary seems to be detrimental to the magnetic moment despite of the formation of $J=1$/2 Ir ions.
Then what is the origin of the observed moment in the slow cooled BYIO-4d?
Since the Curie-Weiss-like paramagnetic behavior is observed by various groups in \BYIO{} in both single crystal and polycrystalline samples, accidental magnetic impurity by itself is unlikely to account for the observed paramagnetism.
While this work does not mean to attribute this solely to antisite disorder or any other mechanism previously proposed by other groups, we want to point out the following:
(1) We would expect some isolated antisite disorder in BYIO-4d.
This kind of ansitite disorder cannot be observed by atomic mapping in STEM study.
The small fraction of the isolated antisite disorder in small scales may not be resolved using x-ray diffraction techniques.
It should be kept in mind that only a small amount of antisite disorder is needed to account for the observed moments.
(2) Lattice defects must be considered when considering possible sources of the observed moments in $J=0$ compounds.
These include nonstoichiometry of both cations and anions, local inhomogeneity, local lattice distortion due to residual stress, foreign inclusions or substitutions by foreign atoms, antisite disorder, twinning and other dislocations.

The sensitivity of magnetism to lattice defects in \BYIO{} resembles that in LaCoO$_3$.
The magnetism and temperature induced spin-state transition in LaCoO$_3$ have attracted attention for 50 years.
A low-spin ($t_{2g}^6e_g^0$, $S=0$) configuration is observed in the ground state due to a larger crystal field splitting of $t_{2g}$ and $e_g$ states than the Hund exchange energy.
However, a nonlinear $M(H)$ curve and a hysteresis loop were observed at low temperatures, and later a ferromagnetic order around 85\,K was reported in small single crystals, nano particles, thin films.\cite{yan2004ferromagnetism,zhou2007ferromagnetism,he2009low,harada2007ferromagnetism,rivadulla2012strain,fuchs2007ferromagnetic}
A large variety of lattice defects has been proposed to account for the observed moment in the $S=0$ state at low temperatures, such as crystal surface, twin boundary, local strain field, oxygen vacancies.
The strong similarity between \BYIO{} and LaCoO$_3$ with respect to the defect induced magnetism suggests that the important effect of lattice defects on local moments is a general phenomena in $J=0$  and $S=0$ compounds.

\section{Summary}

In summary, we investigate the magnetic properties and antisite disorder of three \BYIO{} crystals: (1) a slow cooled crystal, (2) a crystal quenched from 900\degree C, and (3) a crystal that is grown using a faster cooling rate. The scanning transmission electron microscopy imaging show that quenching from 900\degree C introduces antiphase bounary to the crystals and a faster cooling rate during crystal growth leads to clusters of Y and Ir antisite disorder. The magnetic measurements show that \BYIO{} crystals with clusters of antisite disorder have a larger effective moment and a larger saturation moment. The experimental observation indicates the importance of lattice defects in understanding the magnetism in \BYIO{}. Our DFT calculations suggest the magnetic condensation is unlikely as the energy to be gained from superexchange is only one quarter of the spin-orbit gap. However, once Y site is taken by Ir, the enhance superexchange interaction due to increased overlap of orbitals between sites and increased number of orbitals mediating the interactions breaks down the picture of local non-magnetic singlets and thus induces magnetism. Comparison between $J=0$ compound \BYIO{} and $S=0$ compound LaCoO$_3$ suggests that the importnant effect of lattice defects on local moments is a general phenomena in $J=0$  and $S=0$ compounds.

Note: During the preparation of this manuscript, we became aware of the paper by Hammerath et al.\cite{impurity} which also points to the extrinsic origin of the observed magnetism in \BYIO{}.

\section{Acknowledgment}
Work at ORNL was supported by the U.S. Department of Energy, Office of Science, Basic Energy Sciences, Materials Sciences and Engineering Division(QZ, BCS, and JQY). CQ, CS, DM, MR, and NT acknowledge the support of the Center for Emergent Materials, an NSF MRSEC, under Award Number DMR-1420451. DGM acknowledges the support from the Gordon and Betty Moore Foundations EPiQS Initiative through Grant No. GBMF4416. HDZ acknowledges the support from NSF-DMR 1350002. The STEM experiment in this research was conducted at the ORNL’s Center for Nanophase Materials Sciences (CNMS) , which is a DOE Office of Science User Facility.

\appendix

\section{Matrices $\boldsymbol{A}$, $\boldsymbol{B}$, and $\boldsymbol{C}$}
\label{appendixABCMatrices}
The matrices appearing in equation \eqref{equationTripletXYHamiltonian} for $H_{xy}'$ are given explicitly here.
The matrices for $H_{yz}'$ and $H_{zx}'$ may be obtained from cyclic permutations of these matrices.
\begin{widetext}
\begin{equation}
\boldsymbol{A} = \frac{1}{U} \left(
\begin{array}{ccc}
 t_{11}^2+\frac{1}{2} t_{33} t_{11} +\frac{2}{3} t_{12}^2 + \frac{1}{6} t_{33}^2 & t_{11} t_{12} - \frac{1}{6} t_{12} t_{33} & 0 \\
 t_{11} t_{12}-\frac{1}{6} t_{12} t_{33} & t_{11}^2+\frac{1}{2} t_{33} t_{11} +\frac{2}{3} t_{12}^2 +\frac{1}{6} t_{33}^2 & 0 \\
 0 & 0 & \frac{2}{3} t_{11}^2 +\frac{1}{3} t_{33} t_{11} +\frac{2}{3} t_{33}^2 \\
\end{array}
\right)
\end{equation}
\begin{equation}
\boldsymbol{B} = \frac{1}{U} \left(
\begin{array}{ccc}
 -\frac{5}{6} t_{11}^2 -\frac{1}{3}t_{33} t_{11} - \frac{5 }{6}t_{12}^2 -\frac{1}{6} t_{33}^2 & \frac{1}{3} t_{12} t_{33} -t_{11}
   t_{12} & 0 \\
 \frac{1}{3} t_{12} t_{33} - t_{11} t_{12} & -\frac{5}{6} t_{11}^2 -\frac{1}{3}t_{33} t_{11} - \frac{5}{6}   t_{12}^2 -\frac{1}{6} t_{33}^2 & 0 \\
 0 & 0 & -\frac{2}{3} t_{11}^2 -\frac{2}{3} t_{33}^2 \\
\end{array}
\right)
\end{equation}
\begin{equation}
\boldsymbol{C} = \frac{1}{U} \left(
\begin{array}{ccc}
 -\frac{5}{18} t_{11}^2 -\frac{1}{9} t_{33} t_{11} +\frac{1}{6} t_{12}^2 +\frac{1}{18} t_{33}^2 & -\frac{1}{3} t_{11}  t_{12} - \frac{1}{3} t_{33} t_{12} & 0 \\
 -\frac{1}{3} t_{11} t_{12}-\frac{1}{3} t_{33} t_{12} & -\frac{5}{18} t_{11}^2 -\frac{1}{9} t_{33}
   t_{11} +\frac{1}{6} t_{12}^2 +\frac{1}{18} t_{33}^2 & 0 \\
 0 & 0 & \frac{2}{9} t_{11}^2 -\frac{4}{9} t_{33} t_{11} -\frac{1}{9} t_{33}^2 \\
\end{array}
\right)
\end{equation}
\end{widetext}

\section{Inclusion of next-nearest neighbors}
\label{appendixNNNJustification}
Without going through further formalism, we can estimate the effect of including both nearest neighbor (NN) and next-nearest neighbor (NNN) interactions.
Since the NNN Ir ions are in an octahedral configuration around each Ir ion, the NNN case is exactly that of the single perovskite first obtained by Khaliullin.
Since the dispersions of the NN and NNN interactions add linearly for each triplet excitation type, $\omega(\veck{k}) = \omega_\mathrm{NN}(\veck{k}) + \omega_\mathrm{NNN}(\veck{k})$, we can put an upper bound on the result by constructively adding the two energy scales together
\begin{equation}
\frac{16}{3} \frac{t_{33}^2}{U}
+ 10 \frac{t_\mathrm{NNN}^2}{U}
\ge \Delta_\mathrm{SOC}
\label{withNNCondensationRule}
\end{equation}
where the relevant NNN tight-binding parameter obtained from our DFT calculation is $t_\mathrm{NNN} = -18$ meV.
This contributes an extra $2$ meV on the left hand side of \eqref{withNNCondensationRule}, and it only amounts to $47$ meV instead of the original $45$ meV estimate given in the main text.


\begin{thebibliography}{31}%
\makeatletter
\providecommand \@ifxundefined [1]{%
 \@ifx{#1\undefined}
}%
\providecommand \@ifnum [1]{%
 \ifnum #1\expandafter \@firstoftwo
 \else \expandafter \@secondoftwo
 \fi
}%
\providecommand \@ifx [1]{%
 \ifx #1\expandafter \@firstoftwo
 \else \expandafter \@secondoftwo
 \fi
}%
\providecommand \natexlab [1]{#1}%
\providecommand \enquote  [1]{``#1''}%
\providecommand \bibnamefont  [1]{#1}%
\providecommand \bibfnamefont [1]{#1}%
\providecommand \citenamefont [1]{#1}%
\providecommand \href@noop [0]{\@secondoftwo}%
\providecommand \href [0]{\begingroup \@sanitize@url \@href}%
\providecommand \@href[1]{\@@startlink{#1}\@@href}%
\providecommand \@@href[1]{\endgroup#1\@@endlink}%
\providecommand \@sanitize@url [0]{\catcode `\\12\catcode `\$12\catcode
  `\&12\catcode `\#12\catcode `\^12\catcode `\_12\catcode `\%12\relax}%
\providecommand \@@startlink[1]{}%
\providecommand \@@endlink[0]{}%
\providecommand \url  [0]{\begingroup\@sanitize@url \@url }%
\providecommand \@url [1]{\endgroup\@href {#1}{\urlprefix }}%
\providecommand \urlprefix  [0]{URL }%
\providecommand \Eprint [0]{\href }%
\providecommand \doibase [0]{http://dx.doi.org/}%
\providecommand \selectlanguage [0]{\@gobble}%
\providecommand \bibinfo  [0]{\@secondoftwo}%
\providecommand \bibfield  [0]{\@secondoftwo}%
\providecommand \translation [1]{[#1]}%
\providecommand \BibitemOpen [0]{}%
\providecommand \bibitemStop [0]{}%
\providecommand \bibitemNoStop [0]{.\EOS\space}%
\providecommand \EOS [0]{\spacefactor3000\relax}%
\providecommand \BibitemShut  [1]{\csname bibitem#1\endcsname}%
\let\auto@bib@innerbib\@empty
\bibitem [{\citenamefont {Earnshaw}\ \emph {et~al.}(1961)\citenamefont
  {Earnshaw}, \citenamefont {Figgis}, \citenamefont {Lewis},\ and\
  \citenamefont {Peacock}}]{earnshaw1961601}%
  \BibitemOpen
  \bibfield  {author} {\bibinfo {author} {\bibfnamefont {A.}~\bibnamefont
  {Earnshaw}}, \bibinfo {author} {\bibfnamefont {B.}~\bibnamefont {Figgis}},
  \bibinfo {author} {\bibfnamefont {J.}~\bibnamefont {Lewis}}, \ and\ \bibinfo
  {author} {\bibfnamefont {R.}~\bibnamefont {Peacock}},\ }\href@noop {}
  {\bibfield  {journal} {\bibinfo  {journal} {Journal of the Chemical Society
  (Resumed)}\ ,\ \bibinfo {pages} {3132}} (\bibinfo {year} {1961})}\BibitemShut
  {NoStop}%
\bibitem [{\citenamefont {Cao}\ \emph {et~al.}(2014)\citenamefont {Cao},
  \citenamefont {Qi}, \citenamefont {Li}, \citenamefont {Terzic}, \citenamefont
  {Yuan}, \citenamefont {DeLong}, \citenamefont {Murthy},\ and\ \citenamefont
  {Kaul}}]{cao2014novel}%
  \BibitemOpen
  \bibfield  {author} {\bibinfo {author} {\bibfnamefont {G.}~\bibnamefont
  {Cao}}, \bibinfo {author} {\bibfnamefont {T.}~\bibnamefont {Qi}}, \bibinfo
  {author} {\bibfnamefont {L.}~\bibnamefont {Li}}, \bibinfo {author}
  {\bibfnamefont {J.}~\bibnamefont {Terzic}}, \bibinfo {author} {\bibfnamefont
  {S.}~\bibnamefont {Yuan}}, \bibinfo {author} {\bibfnamefont {L.~E.}\
  \bibnamefont {DeLong}}, \bibinfo {author} {\bibfnamefont {G.}~\bibnamefont
  {Murthy}}, \ and\ \bibinfo {author} {\bibfnamefont {R.~K.}\ \bibnamefont
  {Kaul}},\ }\href@noop {} {\bibfield  {journal} {\bibinfo  {journal} {Physical
  review letters}\ }\textbf {\bibinfo {volume} {112}},\ \bibinfo {pages}
  {056402} (\bibinfo {year} {2014})}\BibitemShut {NoStop}%
\bibitem [{\citenamefont {Khaliullin}(2013)}]{khaliullin2013excitonic}%
  \BibitemOpen
  \bibfield  {author} {\bibinfo {author} {\bibfnamefont {G.}~\bibnamefont
  {Khaliullin}},\ }\href@noop {} {\bibfield  {journal} {\bibinfo  {journal}
  {Physical review letters}\ }\textbf {\bibinfo {volume} {111}},\ \bibinfo
  {pages} {197201} (\bibinfo {year} {2013})}\BibitemShut {NoStop}%
\bibitem [{\citenamefont {Bhowal}\ \emph {et~al.}(2015)\citenamefont {Bhowal},
  \citenamefont {Baidya}, \citenamefont {Dasgupta},\ and\ \citenamefont
  {Saha-Dasgupta}}]{bhowal2015breakdown}%
  \BibitemOpen
  \bibfield  {author} {\bibinfo {author} {\bibfnamefont {S.}~\bibnamefont
  {Bhowal}}, \bibinfo {author} {\bibfnamefont {S.}~\bibnamefont {Baidya}},
  \bibinfo {author} {\bibfnamefont {I.}~\bibnamefont {Dasgupta}}, \ and\
  \bibinfo {author} {\bibfnamefont {T.}~\bibnamefont {Saha-Dasgupta}},\
  }\href@noop {} {\bibfield  {journal} {\bibinfo  {journal} {Physical Review
  B}\ }\textbf {\bibinfo {volume} {92}},\ \bibinfo {pages} {121113} (\bibinfo
  {year} {2015})}\BibitemShut {NoStop}%
\bibitem [{\citenamefont {Pajskr}\ \emph {et~al.}(2016)\citenamefont {Pajskr},
  \citenamefont {Nov{\'a}k}, \citenamefont {Pokorn{\`y}}, \citenamefont
  {Koloren{\v{c}}}, \citenamefont {Arita},\ and\ \citenamefont
  {Kune{\v{s}}}}]{pajskr2016possibility}%
  \BibitemOpen
  \bibfield  {author} {\bibinfo {author} {\bibfnamefont {K.}~\bibnamefont
  {Pajskr}}, \bibinfo {author} {\bibfnamefont {P.}~\bibnamefont {Nov{\'a}k}},
  \bibinfo {author} {\bibfnamefont {V.}~\bibnamefont {Pokorn{\`y}}}, \bibinfo
  {author} {\bibfnamefont {J.}~\bibnamefont {Koloren{\v{c}}}}, \bibinfo
  {author} {\bibfnamefont {R.}~\bibnamefont {Arita}}, \ and\ \bibinfo {author}
  {\bibfnamefont {J.}~\bibnamefont {Kune{\v{s}}}},\ }\href@noop {} {\bibfield
  {journal} {\bibinfo  {journal} {Physical Review B}\ }\textbf {\bibinfo
  {volume} {93}},\ \bibinfo {pages} {035129} (\bibinfo {year}
  {2016})}\BibitemShut {NoStop}%
\bibitem [{\citenamefont {Phelan}\ \emph {et~al.}(2016)\citenamefont {Phelan},
  \citenamefont {Seibel}, \citenamefont {Badoe}, \citenamefont {Xie},\ and\
  \citenamefont {Cava}}]{phelan2016influence}%
  \BibitemOpen
  \bibfield  {author} {\bibinfo {author} {\bibfnamefont {B.~F.}\ \bibnamefont
  {Phelan}}, \bibinfo {author} {\bibfnamefont {E.~M.}\ \bibnamefont {Seibel}},
  \bibinfo {author} {\bibfnamefont {D.}~\bibnamefont {Badoe}}, \bibinfo
  {author} {\bibfnamefont {W.}~\bibnamefont {Xie}}, \ and\ \bibinfo {author}
  {\bibfnamefont {R.}~\bibnamefont {Cava}},\ }\href@noop {} {\bibfield
  {journal} {\bibinfo  {journal} {Solid State Communications}\ }\textbf
  {\bibinfo {volume} {236}},\ \bibinfo {pages} {37} (\bibinfo {year}
  {2016})}\BibitemShut {NoStop}%
\bibitem [{\citenamefont {Meetei}\ \emph {et~al.}(2015)\citenamefont {Meetei},
  \citenamefont {Cole}, \citenamefont {Randeria},\ and\ \citenamefont
  {Trivedi}}]{meetei2015novel}%
  \BibitemOpen
  \bibfield  {author} {\bibinfo {author} {\bibfnamefont {O.~N.}\ \bibnamefont
  {Meetei}}, \bibinfo {author} {\bibfnamefont {W.~S.}\ \bibnamefont {Cole}},
  \bibinfo {author} {\bibfnamefont {M.}~\bibnamefont {Randeria}}, \ and\
  \bibinfo {author} {\bibfnamefont {N.}~\bibnamefont {Trivedi}},\ }\href@noop
  {} {\bibfield  {journal} {\bibinfo  {journal} {Physical Review B}\ }\textbf
  {\bibinfo {volume} {91}},\ \bibinfo {pages} {054412} (\bibinfo {year}
  {2015})}\BibitemShut {NoStop}%
\bibitem [{\citenamefont {Sato}\ \emph {et~al.}(2016)\citenamefont {Sato},
  \citenamefont {Shirakawa},\ and\ \citenamefont {Yunoki}}]{sato2016spin}%
  \BibitemOpen
  \bibfield  {author} {\bibinfo {author} {\bibfnamefont {T.}~\bibnamefont
  {Sato}}, \bibinfo {author} {\bibfnamefont {T.}~\bibnamefont {Shirakawa}}, \
  and\ \bibinfo {author} {\bibfnamefont {S.}~\bibnamefont {Yunoki}},\
  }\href@noop {} {\bibfield  {journal} {\bibinfo  {journal} {arXiv preprint
  arXiv:1603.01800}\ } (\bibinfo {year} {2016})}\BibitemShut {NoStop}%
\bibitem [{\citenamefont {Dey}\ \emph {et~al.}(2016)\citenamefont {Dey},
  \citenamefont {Maljuk}, \citenamefont {Efremov}, \citenamefont {Kataeva},
  \citenamefont {Gass}, \citenamefont {Blum}, \citenamefont {Steckel},
  \citenamefont {Gruner}, \citenamefont {Ritschel}, \citenamefont {Wolter}
  \emph {et~al.}}]{dey2016ba}%
  \BibitemOpen
  \bibfield  {author} {\bibinfo {author} {\bibfnamefont {T.}~\bibnamefont
  {Dey}}, \bibinfo {author} {\bibfnamefont {A.}~\bibnamefont {Maljuk}},
  \bibinfo {author} {\bibfnamefont {D.}~\bibnamefont {Efremov}}, \bibinfo
  {author} {\bibfnamefont {O.}~\bibnamefont {Kataeva}}, \bibinfo {author}
  {\bibfnamefont {S.}~\bibnamefont {Gass}}, \bibinfo {author} {\bibfnamefont
  {C.}~\bibnamefont {Blum}}, \bibinfo {author} {\bibfnamefont {F.}~\bibnamefont
  {Steckel}}, \bibinfo {author} {\bibfnamefont {D.}~\bibnamefont {Gruner}},
  \bibinfo {author} {\bibfnamefont {T.}~\bibnamefont {Ritschel}}, \bibinfo
  {author} {\bibfnamefont {A.}~\bibnamefont {Wolter}},  \emph {et~al.},\
  }\href@noop {} {\bibfield  {journal} {\bibinfo  {journal} {Physical Review
  B}\ }\textbf {\bibinfo {volume} {93}},\ \bibinfo {pages} {014434} (\bibinfo
  {year} {2016})}\BibitemShut {NoStop}%
\bibitem [{\citenamefont {Ranjbar}\ \emph {et~al.}(2015)\citenamefont
  {Ranjbar}, \citenamefont {Reynolds}, \citenamefont {Kayser}, \citenamefont
  {Kennedy}, \citenamefont {Hester},\ and\ \citenamefont
  {Kimpton}}]{ranjbar2015structural}%
  \BibitemOpen
  \bibfield  {author} {\bibinfo {author} {\bibfnamefont {B.}~\bibnamefont
  {Ranjbar}}, \bibinfo {author} {\bibfnamefont {E.}~\bibnamefont {Reynolds}},
  \bibinfo {author} {\bibfnamefont {P.}~\bibnamefont {Kayser}}, \bibinfo
  {author} {\bibfnamefont {B.~J.}\ \bibnamefont {Kennedy}}, \bibinfo {author}
  {\bibfnamefont {J.~R.}\ \bibnamefont {Hester}}, \ and\ \bibinfo {author}
  {\bibfnamefont {J.~A.}\ \bibnamefont {Kimpton}},\ }\href@noop {} {\bibfield
  {journal} {\bibinfo  {journal} {Inorganic chemistry}\ }\textbf {\bibinfo
  {volume} {54}},\ \bibinfo {pages} {10468} (\bibinfo {year}
  {2015})}\BibitemShut {NoStop}%
\bibitem [{\citenamefont {Corredor}\ \emph {et~al.}(2016)\citenamefont
  {Corredor}, \citenamefont {Aslan-Cansever}, \citenamefont {Sturza},
  \citenamefont {Manna}, \citenamefont {Maljuk}, \citenamefont {Gass},
  \citenamefont {Zimmermann}, \citenamefont {Dey}, \citenamefont {Blum},
  \citenamefont {Geyer} \emph {et~al.}}]{corredor2016iridium}%
  \BibitemOpen
  \bibfield  {author} {\bibinfo {author} {\bibfnamefont {L.}~\bibnamefont
  {Corredor}}, \bibinfo {author} {\bibfnamefont {G.}~\bibnamefont
  {Aslan-Cansever}}, \bibinfo {author} {\bibfnamefont {M.}~\bibnamefont
  {Sturza}}, \bibinfo {author} {\bibfnamefont {K.}~\bibnamefont {Manna}},
  \bibinfo {author} {\bibfnamefont {A.}~\bibnamefont {Maljuk}}, \bibinfo
  {author} {\bibfnamefont {S.}~\bibnamefont {Gass}}, \bibinfo {author}
  {\bibfnamefont {A.}~\bibnamefont {Zimmermann}}, \bibinfo {author}
  {\bibfnamefont {T.}~\bibnamefont {Dey}}, \bibinfo {author} {\bibfnamefont
  {C.}~\bibnamefont {Blum}}, \bibinfo {author} {\bibfnamefont {M.}~\bibnamefont
  {Geyer}},  \emph {et~al.},\ }\href@noop {} {\bibfield  {journal} {\bibinfo
  {journal} {arXiv preprint arXiv:1606.05104}\ } (\bibinfo {year}
  {2016})}\BibitemShut {NoStop}%
\bibitem [{\citenamefont {Zhao}\ \emph {et~al.}(2016)\citenamefont {Zhao},
  \citenamefont {Calder}, \citenamefont {Aczel}, \citenamefont {McGuire},
  \citenamefont {Sales}, \citenamefont {Mandrus}, \citenamefont {Chen},
  \citenamefont {Trivedi}, \citenamefont {Zhou},\ and\ \citenamefont
  {Yan}}]{zhao2016fragile}%
  \BibitemOpen
  \bibfield  {author} {\bibinfo {author} {\bibfnamefont {Z.}~\bibnamefont
  {Zhao}}, \bibinfo {author} {\bibfnamefont {S.}~\bibnamefont {Calder}},
  \bibinfo {author} {\bibfnamefont {A.}~\bibnamefont {Aczel}}, \bibinfo
  {author} {\bibfnamefont {M.}~\bibnamefont {McGuire}}, \bibinfo {author}
  {\bibfnamefont {B.}~\bibnamefont {Sales}}, \bibinfo {author} {\bibfnamefont
  {D.}~\bibnamefont {Mandrus}}, \bibinfo {author} {\bibfnamefont
  {G.}~\bibnamefont {Chen}}, \bibinfo {author} {\bibfnamefont {N.}~\bibnamefont
  {Trivedi}}, \bibinfo {author} {\bibfnamefont {H.}~\bibnamefont {Zhou}}, \
  and\ \bibinfo {author} {\bibfnamefont {J.-Q.}\ \bibnamefont {Yan}},\
  }\href@noop {} {\bibfield  {journal} {\bibinfo  {journal} {Physical Review
  B}\ }\textbf {\bibinfo {volume} {93}},\ \bibinfo {pages} {134426} (\bibinfo
  {year} {2016})}\BibitemShut {NoStop}%
\bibitem [{\citenamefont {Zhang}\ \emph {et~al.}(2016)\citenamefont {Zhang},
  \citenamefont {Terizc}, \citenamefont {Ye}, \citenamefont {Schlottmann},
  \citenamefont {Zhao}, \citenamefont {Yuan},\ and\ \citenamefont
  {Cao}}]{zhang2016breakdown}%
  \BibitemOpen
  \bibfield  {author} {\bibinfo {author} {\bibfnamefont {H.}~\bibnamefont
  {Zhang}}, \bibinfo {author} {\bibfnamefont {J.}~\bibnamefont {Terizc}},
  \bibinfo {author} {\bibfnamefont {F.}~\bibnamefont {Ye}}, \bibinfo {author}
  {\bibfnamefont {P.}~\bibnamefont {Schlottmann}}, \bibinfo {author}
  {\bibfnamefont {H.}~\bibnamefont {Zhao}}, \bibinfo {author} {\bibfnamefont
  {S.}~\bibnamefont {Yuan}}, \ and\ \bibinfo {author} {\bibfnamefont
  {G.}~\bibnamefont {Cao}},\ }\href@noop {} {\bibfield  {journal} {\bibinfo
  {journal} {arXiv preprint arXiv:1608.07624}\ } (\bibinfo {year}
  {2016})}\BibitemShut {NoStop}%
\bibitem [{\citenamefont {Wang}\ \emph {et~al.}(2014)\citenamefont {Wang},
  \citenamefont {Terzic}, \citenamefont {Qi}, \citenamefont {Ye}, \citenamefont
  {Yuan}, \citenamefont {Aswartham}, \citenamefont {Streltsov}, \citenamefont
  {Khomskii}, \citenamefont {Kaul},\ and\ \citenamefont
  {Cao}}]{wang2014lattice}%
  \BibitemOpen
  \bibfield  {author} {\bibinfo {author} {\bibfnamefont {J.}~\bibnamefont
  {Wang}}, \bibinfo {author} {\bibfnamefont {J.}~\bibnamefont {Terzic}},
  \bibinfo {author} {\bibfnamefont {T.}~\bibnamefont {Qi}}, \bibinfo {author}
  {\bibfnamefont {F.}~\bibnamefont {Ye}}, \bibinfo {author} {\bibfnamefont
  {S.}~\bibnamefont {Yuan}}, \bibinfo {author} {\bibfnamefont {S.}~\bibnamefont
  {Aswartham}}, \bibinfo {author} {\bibfnamefont {S.}~\bibnamefont
  {Streltsov}}, \bibinfo {author} {\bibfnamefont {D.}~\bibnamefont {Khomskii}},
  \bibinfo {author} {\bibfnamefont {R.~K.}\ \bibnamefont {Kaul}}, \ and\
  \bibinfo {author} {\bibfnamefont {G.}~\bibnamefont {Cao}},\ }\href@noop {}
  {\bibfield  {journal} {\bibinfo  {journal} {Physical Review B}\ }\textbf
  {\bibinfo {volume} {90}},\ \bibinfo {pages} {161110} (\bibinfo {year}
  {2014})}\BibitemShut {NoStop}%
\bibitem [{\citenamefont {Laguna-Marco}\ \emph {et~al.}(2015)\citenamefont
  {Laguna-Marco}, \citenamefont {Kayser}, \citenamefont {Alonso}, \citenamefont
  {Mart{\'\i}nez-Lope}, \citenamefont {van Veenendaal}, \citenamefont {Choi},\
  and\ \citenamefont {Haskel}}]{laguna2015electronic}%
  \BibitemOpen
  \bibfield  {author} {\bibinfo {author} {\bibfnamefont {M.}~\bibnamefont
  {Laguna-Marco}}, \bibinfo {author} {\bibfnamefont {P.}~\bibnamefont
  {Kayser}}, \bibinfo {author} {\bibfnamefont {J.}~\bibnamefont {Alonso}},
  \bibinfo {author} {\bibfnamefont {M.}~\bibnamefont {Mart{\'\i}nez-Lope}},
  \bibinfo {author} {\bibfnamefont {M.}~\bibnamefont {van Veenendaal}},
  \bibinfo {author} {\bibfnamefont {Y.}~\bibnamefont {Choi}}, \ and\ \bibinfo
  {author} {\bibfnamefont {D.}~\bibnamefont {Haskel}},\ }\href@noop {}
  {\bibfield  {journal} {\bibinfo  {journal} {Physical Review B}\ }\textbf
  {\bibinfo {volume} {91}},\ \bibinfo {pages} {214433} (\bibinfo {year}
  {2015})}\BibitemShut {NoStop}%
\bibitem [{\citenamefont {Krivanek}\ \emph {et~al.}(2008)\citenamefont
  {Krivanek}, \citenamefont {Corbin}, \citenamefont {Dellby}, \citenamefont
  {Elston}, \citenamefont {Keyse}, \citenamefont {Murfitt}, \citenamefont
  {Own}, \citenamefont {Szilagyi},\ and\ \citenamefont
  {Woodruff}}]{krivanek2008electron}%
  \BibitemOpen
  \bibfield  {author} {\bibinfo {author} {\bibfnamefont {O.}~\bibnamefont
  {Krivanek}}, \bibinfo {author} {\bibfnamefont {G.}~\bibnamefont {Corbin}},
  \bibinfo {author} {\bibfnamefont {N.}~\bibnamefont {Dellby}}, \bibinfo
  {author} {\bibfnamefont {B.}~\bibnamefont {Elston}}, \bibinfo {author}
  {\bibfnamefont {R.}~\bibnamefont {Keyse}}, \bibinfo {author} {\bibfnamefont
  {M.}~\bibnamefont {Murfitt}}, \bibinfo {author} {\bibfnamefont
  {C.}~\bibnamefont {Own}}, \bibinfo {author} {\bibfnamefont {Z.}~\bibnamefont
  {Szilagyi}}, \ and\ \bibinfo {author} {\bibfnamefont {J.}~\bibnamefont
  {Woodruff}},\ }\href@noop {} {\bibfield  {journal} {\bibinfo  {journal}
  {Ultramicroscopy}\ }\textbf {\bibinfo {volume} {108}},\ \bibinfo {pages}
  {179} (\bibinfo {year} {2008})}\BibitemShut {NoStop}%
\bibitem [{\citenamefont {Esser}\ \emph {et~al.}(2016)\citenamefont {Esser},
  \citenamefont {Hauser}, \citenamefont {Williams}, \citenamefont {Allen},
  \citenamefont {Woodward}, \citenamefont {Yang},\ and\ \citenamefont
  {McComb}}]{esser2016quantitative}%
  \BibitemOpen
  \bibfield  {author} {\bibinfo {author} {\bibfnamefont {B.}~\bibnamefont
  {Esser}}, \bibinfo {author} {\bibfnamefont {A.}~\bibnamefont {Hauser}},
  \bibinfo {author} {\bibfnamefont {R.}~\bibnamefont {Williams}}, \bibinfo
  {author} {\bibfnamefont {L.}~\bibnamefont {Allen}}, \bibinfo {author}
  {\bibfnamefont {P.}~\bibnamefont {Woodward}}, \bibinfo {author}
  {\bibfnamefont {F.}~\bibnamefont {Yang}}, \ and\ \bibinfo {author}
  {\bibfnamefont {D.}~\bibnamefont {McComb}},\ }\href@noop {} {\bibfield
  {journal} {\bibinfo  {journal} {Physical review letters}\ }\textbf {\bibinfo
  {volume} {117}},\ \bibinfo {pages} {176101} (\bibinfo {year}
  {2016})}\BibitemShut {NoStop}%
\bibitem [{\citenamefont {Yan}\ \emph {et~al.}(2004{\natexlab{a}})\citenamefont
  {Yan}, \citenamefont {Zhou},\ and\ \citenamefont {Goodenough}}]{yan2004bond}%
  \BibitemOpen
  \bibfield  {author} {\bibinfo {author} {\bibfnamefont {J.-Q.}\ \bibnamefont
  {Yan}}, \bibinfo {author} {\bibfnamefont {J.-S.}\ \bibnamefont {Zhou}}, \
  and\ \bibinfo {author} {\bibfnamefont {J.}~\bibnamefont {Goodenough}},\
  }\href@noop {} {\bibfield  {journal} {\bibinfo  {journal} {Physical Review
  B}\ }\textbf {\bibinfo {volume} {69}},\ \bibinfo {pages} {134409} (\bibinfo
  {year} {2004}{\natexlab{a}})}\BibitemShut {NoStop}%
\bibitem [{RIX()}]{RIXS}%
  \BibitemOpen
  \href@noop {} {}\bibinfo {note} {{European Synchrotron Radiation Facility
  experimental report, Resonant inelastic x-ray scattering of the new iridium
  compound Ba2YIrO6.2015.}}\BibitemShut {Stop}%
\bibitem [{\citenamefont {Blaha}\ \emph {et~al.}(2001)\citenamefont {Blaha},
  \citenamefont {Schwarz}, \citenamefont {Madsen}, \citenamefont {Kvasnicka},\
  and\ \citenamefont {Luitz}}]{blaha2001wien2k}%
  \BibitemOpen
  \bibfield  {author} {\bibinfo {author} {\bibfnamefont {P.}~\bibnamefont
  {Blaha}}, \bibinfo {author} {\bibfnamefont {K.}~\bibnamefont {Schwarz}},
  \bibinfo {author} {\bibfnamefont {G.}~\bibnamefont {Madsen}}, \bibinfo
  {author} {\bibfnamefont {D.}~\bibnamefont {Kvasnicka}}, \ and\ \bibinfo
  {author} {\bibfnamefont {J.}~\bibnamefont {Luitz}},\ }\href@noop {}
  {\bibfield  {journal} {\bibinfo  {journal} {An augmented plane wave+ local
  orbitals program for calculating crystal properties}\ } (\bibinfo {year}
  {2001})}\BibitemShut {NoStop}%
\bibitem [{\citenamefont {Mostofi}\ \emph {et~al.}(2014)\citenamefont
  {Mostofi}, \citenamefont {Yates}, \citenamefont {Pizzi}, \citenamefont {Lee},
  \citenamefont {Souza}, \citenamefont {Vanderbilt},\ and\ \citenamefont
  {Marzari}}]{mostofi2014updated}%
  \BibitemOpen
  \bibfield  {author} {\bibinfo {author} {\bibfnamefont {A.~A.}\ \bibnamefont
  {Mostofi}}, \bibinfo {author} {\bibfnamefont {J.~R.}\ \bibnamefont {Yates}},
  \bibinfo {author} {\bibfnamefont {G.}~\bibnamefont {Pizzi}}, \bibinfo
  {author} {\bibfnamefont {Y.-S.}\ \bibnamefont {Lee}}, \bibinfo {author}
  {\bibfnamefont {I.}~\bibnamefont {Souza}}, \bibinfo {author} {\bibfnamefont
  {D.}~\bibnamefont {Vanderbilt}}, \ and\ \bibinfo {author} {\bibfnamefont
  {N.}~\bibnamefont {Marzari}},\ }\href@noop {} {\bibfield  {journal} {\bibinfo
   {journal} {Computer Physics Communications}\ }\textbf {\bibinfo {volume}
  {185}},\ \bibinfo {pages} {2309} (\bibinfo {year} {2014})}\BibitemShut
  {NoStop}%
\bibitem [{\citenamefont {Lim}\ \emph {et~al.}(2016)\citenamefont {Lim},
  \citenamefont {Kim}, \citenamefont {Sung}, \citenamefont {Rhyim},
  \citenamefont {Jeen}, \citenamefont {Yun}, \citenamefont {Kim}, \citenamefont
  {Song}, \citenamefont {Lee}, \citenamefont {Chung} \emph
  {et~al.}}]{lim2016insights}%
  \BibitemOpen
  \bibfield  {author} {\bibinfo {author} {\bibfnamefont {T.-W.}\ \bibnamefont
  {Lim}}, \bibinfo {author} {\bibfnamefont {S.-D.}\ \bibnamefont {Kim}},
  \bibinfo {author} {\bibfnamefont {K.-D.}\ \bibnamefont {Sung}}, \bibinfo
  {author} {\bibfnamefont {Y.-M.}\ \bibnamefont {Rhyim}}, \bibinfo {author}
  {\bibfnamefont {H.}~\bibnamefont {Jeen}}, \bibinfo {author} {\bibfnamefont
  {J.}~\bibnamefont {Yun}}, \bibinfo {author} {\bibfnamefont {K.-H.}\
  \bibnamefont {Kim}}, \bibinfo {author} {\bibfnamefont {K.-M.}\ \bibnamefont
  {Song}}, \bibinfo {author} {\bibfnamefont {S.}~\bibnamefont {Lee}}, \bibinfo
  {author} {\bibfnamefont {S.-Y.}\ \bibnamefont {Chung}},  \emph {et~al.},\
  }\href@noop {} {\bibfield  {journal} {\bibinfo  {journal} {Scientific
  reports}\ }\textbf {\bibinfo {volume} {6}},\ \bibinfo {pages} {19746}
  (\bibinfo {year} {2016})}\BibitemShut {NoStop}%
\bibitem [{\citenamefont {Navarro}\ \emph {et~al.}(2001)\citenamefont
  {Navarro}, \citenamefont {Bibes}, \citenamefont {Roig}, \citenamefont
  {Martinez},\ and\ \citenamefont {Fontcuberta}}]{navarro2001cationic}%
  \BibitemOpen
  \bibfield  {author} {\bibinfo {author} {\bibfnamefont {J.}~\bibnamefont
  {Navarro}}, \bibinfo {author} {\bibfnamefont {M.}~\bibnamefont {Bibes}},
  \bibinfo {author} {\bibfnamefont {A.}~\bibnamefont {Roig}}, \bibinfo {author}
  {\bibfnamefont {B.}~\bibnamefont {Martinez}}, \ and\ \bibinfo {author}
  {\bibfnamefont {J.}~\bibnamefont {Fontcuberta}},\ }\href@noop {} {\bibfield
  {journal} {\bibinfo  {journal} {Applied Physics Letters}\ }\textbf {\bibinfo
  {volume} {78}},\ \bibinfo {pages} {781} (\bibinfo {year} {2001})}\BibitemShut
  {NoStop}%
\bibitem [{\citenamefont {Dey}\ \emph {et~al.}(2013)\citenamefont {Dey},
  \citenamefont {Mahajan}, \citenamefont {Kumar}, \citenamefont {Koteswararao},
  \citenamefont {Chou}, \citenamefont {Omrani},\ and\ \citenamefont
  {Ronnow}}]{dey2013possible}%
  \BibitemOpen
  \bibfield  {author} {\bibinfo {author} {\bibfnamefont {T.}~\bibnamefont
  {Dey}}, \bibinfo {author} {\bibfnamefont {A.}~\bibnamefont {Mahajan}},
  \bibinfo {author} {\bibfnamefont {R.}~\bibnamefont {Kumar}}, \bibinfo
  {author} {\bibfnamefont {B.}~\bibnamefont {Koteswararao}}, \bibinfo {author}
  {\bibfnamefont {F.}~\bibnamefont {Chou}}, \bibinfo {author} {\bibfnamefont
  {A.}~\bibnamefont {Omrani}}, \ and\ \bibinfo {author} {\bibfnamefont
  {H.}~\bibnamefont {Ronnow}},\ }\href@noop {} {\bibfield  {journal} {\bibinfo
  {journal} {Physical Review B}\ }\textbf {\bibinfo {volume} {88}},\ \bibinfo
  {pages} {134425} (\bibinfo {year} {2013})}\BibitemShut {NoStop}%
\bibitem [{\citenamefont {Panda}\ \emph {et~al.}(2015)\citenamefont {Panda},
  \citenamefont {Bhowal}, \citenamefont {Li}, \citenamefont {Ganguly},
  \citenamefont {Valent{\'\i}}, \citenamefont {Nordstr{\"o}m},\ and\
  \citenamefont {Dasgupta}}]{panda2015electronic}%
  \BibitemOpen
  \bibfield  {author} {\bibinfo {author} {\bibfnamefont {S.}~\bibnamefont
  {Panda}}, \bibinfo {author} {\bibfnamefont {S.}~\bibnamefont {Bhowal}},
  \bibinfo {author} {\bibfnamefont {Y.}~\bibnamefont {Li}}, \bibinfo {author}
  {\bibfnamefont {S.}~\bibnamefont {Ganguly}}, \bibinfo {author} {\bibfnamefont
  {R.}~\bibnamefont {Valent{\'\i}}}, \bibinfo {author} {\bibfnamefont
  {L.}~\bibnamefont {Nordstr{\"o}m}}, \ and\ \bibinfo {author} {\bibfnamefont
  {I.}~\bibnamefont {Dasgupta}},\ }\href@noop {} {\bibfield  {journal}
  {\bibinfo  {journal} {Physical Review B}\ }\textbf {\bibinfo {volume} {92}},\
  \bibinfo {pages} {180403} (\bibinfo {year} {2015})}\BibitemShut {NoStop}%
\bibitem [{\citenamefont {Yan}\ \emph {et~al.}(2004{\natexlab{b}})\citenamefont
  {Yan}, \citenamefont {Zhou},\ and\ \citenamefont
  {Goodenough}}]{yan2004ferromagnetism}%
  \BibitemOpen
  \bibfield  {author} {\bibinfo {author} {\bibfnamefont {J.-Q.}\ \bibnamefont
  {Yan}}, \bibinfo {author} {\bibfnamefont {J.-S.}\ \bibnamefont {Zhou}}, \
  and\ \bibinfo {author} {\bibfnamefont {J.}~\bibnamefont {Goodenough}},\
  }\href@noop {} {\bibfield  {journal} {\bibinfo  {journal} {Physical Review
  B}\ }\textbf {\bibinfo {volume} {70}},\ \bibinfo {pages} {014402} (\bibinfo
  {year} {2004}{\natexlab{b}})}\BibitemShut {NoStop}%
\bibitem [{\citenamefont {Zhou}\ \emph {et~al.}(2007)\citenamefont {Zhou},
  \citenamefont {Shi}, \citenamefont {Zhao}, \citenamefont {He}, \citenamefont
  {Yang},\ and\ \citenamefont {Zhang}}]{zhou2007ferromagnetism}%
  \BibitemOpen
  \bibfield  {author} {\bibinfo {author} {\bibfnamefont {S.}~\bibnamefont
  {Zhou}}, \bibinfo {author} {\bibfnamefont {L.}~\bibnamefont {Shi}}, \bibinfo
  {author} {\bibfnamefont {J.}~\bibnamefont {Zhao}}, \bibinfo {author}
  {\bibfnamefont {L.}~\bibnamefont {He}}, \bibinfo {author} {\bibfnamefont
  {H.}~\bibnamefont {Yang}}, \ and\ \bibinfo {author} {\bibfnamefont
  {S.}~\bibnamefont {Zhang}},\ }\href@noop {} {\bibfield  {journal} {\bibinfo
  {journal} {Physical Review B}\ }\textbf {\bibinfo {volume} {76}},\ \bibinfo
  {pages} {172407} (\bibinfo {year} {2007})}\BibitemShut {NoStop}%
\bibitem [{\citenamefont {He}\ \emph {et~al.}(2009)\citenamefont {He},
  \citenamefont {Zheng}, \citenamefont {Mitchell}, \citenamefont {Foo},
  \citenamefont {Cava},\ and\ \citenamefont {Leighton}}]{he2009low}%
  \BibitemOpen
  \bibfield  {author} {\bibinfo {author} {\bibfnamefont {C.}~\bibnamefont
  {He}}, \bibinfo {author} {\bibfnamefont {H.}~\bibnamefont {Zheng}}, \bibinfo
  {author} {\bibfnamefont {J.}~\bibnamefont {Mitchell}}, \bibinfo {author}
  {\bibfnamefont {M.}~\bibnamefont {Foo}}, \bibinfo {author} {\bibfnamefont
  {R.}~\bibnamefont {Cava}}, \ and\ \bibinfo {author} {\bibfnamefont
  {C.}~\bibnamefont {Leighton}},\ }\href@noop {} {\bibfield  {journal}
  {\bibinfo  {journal} {Applied Physics Letters}\ }\textbf {\bibinfo {volume}
  {94}},\ \bibinfo {pages} {102514} (\bibinfo {year} {2009})}\BibitemShut
  {NoStop}%
\bibitem [{\citenamefont {Harada}\ \emph {et~al.}(2007)\citenamefont {Harada},
  \citenamefont {Taniyama}, \citenamefont {Takeuchi}, \citenamefont {Sato},
  \citenamefont {Kyomen},\ and\ \citenamefont
  {Itoh}}]{harada2007ferromagnetism}%
  \BibitemOpen
  \bibfield  {author} {\bibinfo {author} {\bibfnamefont {A.}~\bibnamefont
  {Harada}}, \bibinfo {author} {\bibfnamefont {T.}~\bibnamefont {Taniyama}},
  \bibinfo {author} {\bibfnamefont {Y.}~\bibnamefont {Takeuchi}}, \bibinfo
  {author} {\bibfnamefont {T.}~\bibnamefont {Sato}}, \bibinfo {author}
  {\bibfnamefont {T.}~\bibnamefont {Kyomen}}, \ and\ \bibinfo {author}
  {\bibfnamefont {M.}~\bibnamefont {Itoh}},\ }\href@noop {} {\bibfield
  {journal} {\bibinfo  {journal} {Physical Review B}\ }\textbf {\bibinfo
  {volume} {75}},\ \bibinfo {pages} {184426} (\bibinfo {year}
  {2007})}\BibitemShut {NoStop}%
\bibitem [{\citenamefont {Rivadulla}\ \emph {et~al.}(2012)\citenamefont
  {Rivadulla}, \citenamefont {Bi}, \citenamefont {Bauer}, \citenamefont
  {Rivas-Murias}, \citenamefont {Vila-Fungueiriño},\ and\ \citenamefont
  {Jia}}]{rivadulla2012strain}%
  \BibitemOpen
  \bibfield  {author} {\bibinfo {author} {\bibfnamefont {F.}~\bibnamefont
  {Rivadulla}}, \bibinfo {author} {\bibfnamefont {Z.}~\bibnamefont {Bi}},
  \bibinfo {author} {\bibfnamefont {E.}~\bibnamefont {Bauer}}, \bibinfo
  {author} {\bibfnamefont {B.}~\bibnamefont {Rivas-Murias}}, \bibinfo {author}
  {\bibfnamefont {J.~M.}\ \bibnamefont {Vila-Fungueiriño}}, \ and\ \bibinfo
  {author} {\bibfnamefont {Q.}~\bibnamefont {Jia}},\ }\href@noop {} {\bibfield
  {journal} {\bibinfo  {journal} {Chemistry of Materials}\ }\textbf {\bibinfo
  {volume} {25}},\ \bibinfo {pages} {55} (\bibinfo {year} {2012})}\BibitemShut
  {NoStop}%
\bibitem [{\citenamefont {Fuchs}\ \emph {et~al.}(2007)\citenamefont {Fuchs},
  \citenamefont {Pinta}, \citenamefont {Schwarz}, \citenamefont {Schweiss},
  \citenamefont {Nagel}, \citenamefont {Schuppler}, \citenamefont {Schneider},
  \citenamefont {Merz}, \citenamefont {Roth},\ and\ \citenamefont
  {L{\"o}hneysen}}]{fuchs2007ferromagnetic}%
  \BibitemOpen
  \bibfield  {author} {\bibinfo {author} {\bibfnamefont {D.}~\bibnamefont
  {Fuchs}}, \bibinfo {author} {\bibfnamefont {C.}~\bibnamefont {Pinta}},
  \bibinfo {author} {\bibfnamefont {T.}~\bibnamefont {Schwarz}}, \bibinfo
  {author} {\bibfnamefont {P.}~\bibnamefont {Schweiss}}, \bibinfo {author}
  {\bibfnamefont {P.}~\bibnamefont {Nagel}}, \bibinfo {author} {\bibfnamefont
  {S.}~\bibnamefont {Schuppler}}, \bibinfo {author} {\bibfnamefont
  {R.}~\bibnamefont {Schneider}}, \bibinfo {author} {\bibfnamefont
  {M.}~\bibnamefont {Merz}}, \bibinfo {author} {\bibfnamefont {G.}~\bibnamefont
  {Roth}}, \ and\ \bibinfo {author} {\bibfnamefont {H.~v.}\ \bibnamefont
  {L{\"o}hneysen}},\ }\href@noop {} {\bibfield  {journal} {\bibinfo  {journal}
  {Physical Review B}\ }\textbf {\bibinfo {volume} {75}},\ \bibinfo {pages}
  {144402} (\bibinfo {year} {2007})}\BibitemShut {NoStop}%
\bibitem [{imp()}]{impurity}%
  \BibitemOpen
  \href@noop {} {}\bibinfo {note} {{F. Hammerath, R. Sarkar, S. Kamusella, C.
  Baines, H.-H. Klauss, T. Dey, A. Maljuk, S. Gaß, A.U.B. Wolter, H.-J. Grafe,
  S. Wurmehl, B. Büchner. Diluted paramagnetic impurities in nonmagnetic
  Ba2YIrO6. arxiv:1706.06027}}\BibitemShut {NoStop}%
\end{thebibliography}

%

\end{document}